%% file: review.tex
\documentclass{ar-1col}
\usepackage[numbers]{natbib}
\usepackage{url}
\usepackage{subfigure}
\usepackage{amssymb}
\usepackage{mathtools}

\jname{Ann. Rev. Nucl. Part. Sci.}
\jyear{2022}
\begin{document}

% Page header
\markboth{Kurahashi, Murase \& Santander}{High-Energy Extragalactic Neutrino Astrophysics}

% Title
\title{High-Energy Extragalactic Neutrino Astrophysics}

%Authors, affiliations address.
\author{Naoko Kurahashi,$^1$, Kohta Murase$^{2,3}$, and Marcos Santander$^4$
\affil{$^1$Department of Physics, Drexel University, Philadelphia, PA, USA, 19104; email: naoko@drexel.edu}
\affil{$^2$Department of Physics, Department of Astronomy \& Astrophysics, Institute for Gravitation and the Cosmos, The Pennsylvania State University, University Park, PA 16802, USA; 
email: murase@psu.edu}
\affil{$^3$Center for Gravitational Physics, Yukawa Institute for Theoretical Physics, Kyoto, Kyoto 606-8502, Japan}
\affil{$^4$Department of Physics and Astronomy, University of Alabama, Tuscaloosa, AL 35487, USA; email: jmsantander@ua.edu}
}

%Abstract
\begin{abstract}
The detection of an astrophysical flux of neutrinos in the TeV-PeV energy range by the IceCube observatory has opened new possibilities for the study of extreme cosmic accelerators. The apparent isotropy of the neutrino arrival directions favors an extragalactic origin for this flux, potentially created by a large population of distant sources. Recent evidence for the detection of neutrino emission from extragalactic sources include the active galaxies TXS 0506+056 and NGC 1068.

We here review the current status of the search for the sources of the high-energy neutrino flux, concentrating on its extragalactic contribution. We discuss the implications of these observations for multimessenger studies of cosmic sources and present an outlook for how additional observations by current and future instruments will help address fundamental questions in the emerging field of high-energy neutrino astronomy.

%this is the old one from last year:
%The detection of an astrophysical flux of neutrinos in the TeV-PeV energy range by the IceCube observatory has opened new possibilities for the study of extreme cosmic accelerators. The apparent isotropy of the neutrino arrival directions favors an extragalactic origin for this flux, potentially created by a large population of distant sources. Recent evidence for the detection of neutrino emission from the active galaxy TXS 0506+056 could also signify the start of time-domain neutrino astrophysics.

%We review the current status of the search for the sources of the high-energy neutrino flux, concentrating on its extragalactic contribution. We discuss the implications of these observations for multimessenger studies of cosmic sources and present an outlook for how additional observations by current and future instruments will help address fundamental questions in the emerging field of high-energy neutrino astronomy.

\end{abstract}

%Keywords, etc.
\maketitle

%Table of Contents
\tableofcontents

% Intro section 
\input{intro.tex}

% Telescope section
\input{nutels.tex}

% Diffuse section
\input{diffuse.tex}

% Exp results
\input{xgal_exp.tex}

% Implications for sources 
\input{xgal_srcs.tex}

% Prospects for neutrino studies
%\input{prospects.tex}

% Bigger MMA picture
\input{mma_picture.tex}

% Future Issues
\begin{issues}[FUTURE ISSUES]
\begin{enumerate}
\item Fully enabling extragalactic neutrino astronomy is within reach of the next generation of high-energy neutrino telescopes.

\item The sensitivity of these studies will strongly rely on improvements on sensitivity and angular resolution. 

\item Upcoming electromagnetic instruments will play a critical role in pinpointing multimessenger counterparts to neutrino signals and to characterize their properties once sources are strongly detected. Joint GW-neutrino studies could also deliver multimessenger detections.

\item The interconnection of neutrino telescopes, both among themselves and with electromagnetic/gravitational-wave facilities, will be key in identifying transient or bursting/flaring neutrino sources.

\item Further theoretical investigations, including numerical simulations on source dynamics and particle acceleration mechanisms, will be necessary to establish concordance pictures of multimessenger emission and understand physics of the neutrino sources. 

\end{enumerate}
\end{issues}

%Disclosure
\section*{DISCLOSURE STATEMENT}
%If the authors have noting to disclose, the following statement will be used: 
The authors are not aware of any affiliations, memberships, funding, or financial holdings that
might be perceived as affecting the objectivity of this review. 

% Acknowledgements
\section*{ACKNOWLEDGMENTS}
We thank Theo Glauch for useful comments. 
N.K.N. is supported by NSF Grant PHYS-1847827. 
The work of K.M. is supported by the NSF Grant No.~AST-1908689, No.~AST-2108466 and No.~AST-2108467, and KAKENHI No.~20H01901 and No.~20H05852. 
M.S. is supported by NSF grants PHY-1914579, PHY-1913607, PHY-1828168, PHY-2012944, OIA-2019597, AST-2108517 and NASA grants 80NSSC20K0049, 80NSSC20K0473, 80NSSC20K1494, and 80NSSC20K1587. 
In memory of Tom Gaisser, who was our kind mentor and cornerstone of our field.

% References
\bibliographystyle{ar-style5}
\bibliography{review,kmurase}

\end{document}

%% file: intro.tex
% Heading 1
\section{INTRODUCTION \label{sec:intro}} % 1 page 
Neutrinos are a unique probe of the high-energy cosmos. Their small interaction cross sections allow them to escape the dense regions of astrophysical sources that may be opaque to photons, and their paths are unaffected by intervening electromagnetic fields as they are electrically neutral. Neutrinos have been used as astrophysical messengers for decades, with the detection of MeV neutrinos from the Sun~\cite{PhysRevLett.20.1205, PhysRevLett.20.1209} and supernova 1987A~\cite{PhysRevLett.58.1490, PhysRevLett.58.1494}.

At higher energies, in the GeV to TeV range and above, astrophysical neutrinos are expected to be produced in the interactions of hadronic particles, either at their acceleration site or during propagation through interstellar or intergalactic space. Neutrinos may therefore trace astrophysical particle acceleration sites and provide the solution to the cosmic-ray origin mystery. The discovery of an astrophysical neutrino flux in the 10 TeV - 10 PeV range~\cite{Aartsen:2013jdh} represents a breakthrough towards enabling high-energy neutrino astronomy. 
With no strong anisotropy observed in the arrival direction of these astrophysical neutrinos, especially lacking any directional signature that follows the Galactic Plane, it is likely that this flux is dominated by extragalactic sources. The observation of the high-energy diffuse flux combined with the tantalizing evidence for neutrino emission from the active galaxies TXS 0506+056 and NGC 1068 has put us on the doorsteps of extragalactic neutrino astronomy. 

% Missing UHECR connection, potential sources. Production and gamma connection.

We review here recent results from the study of high-energy astrophysical neutrinos and its implications for the sources of this flux, in particular the potential extragalactic contribution. We describe the current experimental landscape and introduce the detectors coming online in the near future, as well as discuss the connection between the neutrino observations and those in a broader multimessenger context involving photons, cosmic rays, and gravitational waves. 

For the purpose of this review, we define high-energy neutrinos as those with energies starting in the GeV to TeV range and above. The reader is referred to recent reviews on the detection of MeV neutrinos from the Sun~\cite{doi:10.1146/annurev-astro-081811-125539} and from supernovae~\cite{doi:10.1146/annurev-nucl-102711-095006, nusn_tamborra} that gave birth to neutrino astronomy as a field.

%Extragalactic neutrino astrophysics has become a reality with two important breakthroughs in the last decade. First is the observations of a diffuse astrophysical flux of neutrinos in tens of TeV to PeVs.  The other is the first indication of a high-energy neutrino source, the active galaxy TXS0506+056~\cite{}. %While the emission mechanisms of high-energy neutrinos and their relation with gamma-ray emission in a flaring blazar is still under debate, this observation marks the dawn of multi-messenger extragalactic neutrino astrophysics. 

%Celestial radiation of neutrinos have been measured for decades. The first astrophysical source to be detected in neutrinos was the Sun in the 1950s. Neutrinos with energies as low as several hundred keV to hundreds of MeVs produced in nuclear fusion reactions from the Sun's core have since been detected. Another celestial source observed in MeV neutrinos is a supernova, SN1987A, in the Large Magellanic Cloud. At higher energies, extragalactic neutrino emissions are expected to dominate.  Very brief history of neutrino astronomy (sun, SN1987A)
%Very brief overview of energy ranges (solar to GZK), define our energy of interest, potential sources
%Quick review of counterparts: UHECR and gammas

%% file: nutels.tex
\section{NEUTRINO OBSERVATORIES, PRESENT AND FUTURE \label{sec:nutel}} % 1 page

The extremely small interaction cross sections ($\sigma_{\nu p} \sim 10^{-38}~{(E_{\nu}/{\rm GeV})}$~cm$^{2}$) of neutrinos present a challenge to their detection upon arrival at Earth. Therefore, neutrino detectors have historically been large. As the neutrino energy increases, so does its cross section, but celestial neutrino emissions follow steeply falling power laws for which the rising cross section cannot compensate. The end result is a larger detector to target higher energy neutrinos. Early estimates of the astrophysical neutrino flux level in the TeV-PeV energy range pointed to the necessity of a kilometer-scale neutrino detector, which would require a naturally-occurring detection medium to make its construction economically feasible. 

Current high-energy neutrino detectors implement the ``water Cherenkov'' technique originally proposed in 1960 by Markov~\cite{Markov:1960vja}, and independently by Reines~\cite{Reines:1960we} and Greisen~\cite{Greisen:1960wc}. 
In this approach, a large natural body of water such as a deep lake, sea or glacial ice is instrumented with a volumetric array of light sensors such as photomultiplier tubes (PMTs). These PMTs detect the Cherenkov photons emitted by relativistic charged particles produced in neutrino interactions which allow the energy, direction and flavor of the neutrino to be inferred. 
The first proposal for a kilometer-scale neutrino detector instrumented with PMTs was put forward by the DUMAND Collaboration~\cite{RevModPhys.64.259} followed by the parallel development of the Baikal, AMANDA and ANTARES instruments. The history of the pioneering efforts to make high-energy neutrino astronomy possible is recounted in ~\cite{2012EPJH...37..515S}.

\begin{figure}[h]
\includegraphics[width=\linewidth]{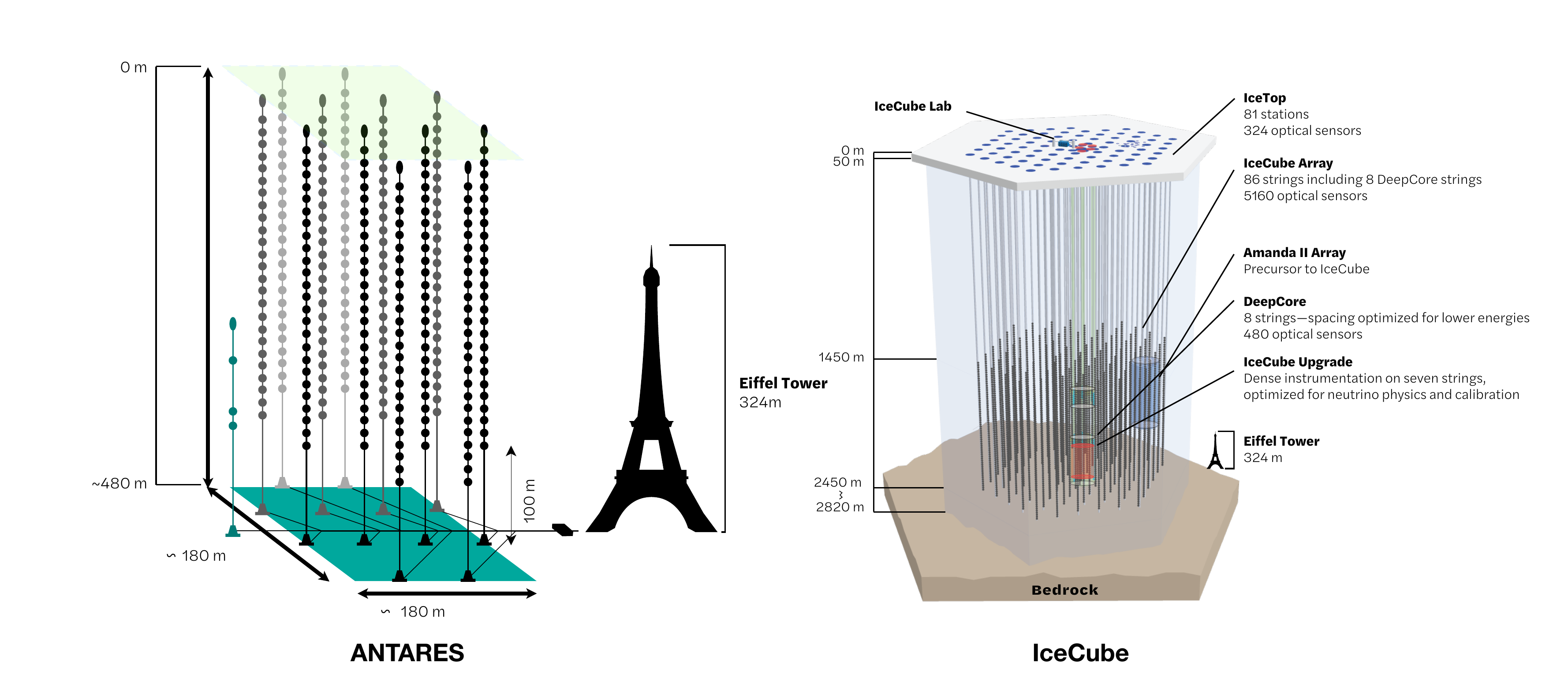}
%\subfigure{\includegraphics[width=2in]{figs/ANTARES_detector.png}}
%\subfigure{\includegraphics[width=2in]{figs/ICECUBE_detector.png}}
\caption{The current generation of high-energy neutrino telescopes: ANTARES and IceCube. The Eiffel tower is shown for scale next to both detectors as their scale is different.}
\label{fig:detectors}
\end{figure}

The two fully-commissioned neutrino telescopes currently in operation are ANTARES and IceCube, shown in \textbf{Figure~\ref{fig:detectors}}.
ANTARES~\cite{Collaboration:2011nsa} has been operating in its final configuration since 2008, with 885 PMTs instrumenting a volume of about 0.01 km$^{3}$ of Mediterranean sea off the coast of Marseille, France. ANTARES will be superseded by KM3NeT~\cite{Adrian-Martinez:2016fdl}, a neutrino telescope with two main components: ORCA, dedicated to the study of neutrino properties, and ARCA, optimized for high-energy neutrino astrophysics. In its final configuration, ARCA will consist of two blocks, each with 115 detector units (or ``strings'') with 18 optical modules per string, which combined will instrument a volume of $\sim 1$ km$^{3}$. As of 2021, six detector units were operational in the ORCA site (off the coast of Toulon, France), and six were operational at the ARCA site near Portopalo di Capo Passero, Sicily, Italy~\cite{Sinopoulou:2021rgv}.

GVD (Gigaton Volume Detector)~\cite{Avrorin:2019dli} is a current effort to build a km$^{3}$-scale detector in Lake Baikal, Russia, following the operation of previous detectors at the site. The detector consists of ``clusters'' of 288 PMTs arranged along eight ``strings'' each. The first phase, GVD-1, was deployed in 2021 and consists of eight clusters with 2304 PMTs covering a volume of 0.4 km$^{3}$~\cite{Baikal-GVD:2021zsq}.

IceCube~\cite{Aartsen:2016nxy, doi:10.1146/annurev-nucl-102313-025321}, deployed in the deep Antarctic ice near the South Pole, is the most sensitive high-energy neutrino telescope currently in operation and the first to reach the km$^{3}$-scale. The detector was completed in 2010 and consists of 5,160 PMTs distributed over 86 strings. The IceCube observatory includes IceTop~\cite{2013NIMPA.700..188A}, an air-shower array on the ice surface for cosmic-ray studies, and DeepCore, a denser, deeper array with improved sensitivity to neutrinos with energies down to 10 GeV~\cite{Collaboration:2011ym}. Most of the results we will discuss in this review come from IceCube, as the preeminent neutrino observatory of the current generation.

The capability of these detectors to determine the energy, incoming direction and flavor of the incoming neutrinos relies on the physics of weak interactions and the energy losses of the secondary particles they produce. Neutrinos and anit-neutrinos are almost always indistinguishable, and interchangeable in this discussion. In the TeV range and above, neutrino interactions are dominated by deep inelastic scatterings between the incoming neutrino and quarks in the nucleon of the detection medium~\cite{2019inelasticity}. Charged-current (CC) interactions mediated by the exchange of a W$^{\pm}$ boson result in a particle shower accompanied by a charged lepton with the same flavor as the incoming neutrino (i.e. $\nu_{\ell} + N \rightarrow \ell^{-} + X$). In a neutral-current (NC) interaction, $\nu_{\ell} + N \rightarrow \nu_{\ell} + X$, a particle shower is produced as the neutrino transfers a fraction of its energy to the nucleon via the exchange of a $Z^{0}$ boson. An additional channel is associated with the resonant production of $W^{-}$ in $\bar{\nu}_{e} e^{-}$ interactions, the Glashow resonance~\cite{PhysRev.118.316}, which dominates over neutrino-nucleon interactions at a neutrino energy of 6.3 PeV.

The main tool for high-energy neutrino astronomy is the detection of $\nu_{\mu}+\bar{\nu}_{\mu}$ via CC interactions. The low energy-loss rate of muons, with a longer lifetime compared to taus, allows them to travel for kilometers in water or ice while emitting Cherenkov photons~\cite{Chirkin:2004hz}. This has a two-fold effect: it increases the effective size of the detector, since it is now sensitive to muons from neutrinos interacting outside the instrumented volume, and it provides good angular resolution, as the long muon ``tracks'' provide a long lever arm for muon directional reconstructions. The kinematic opening angle between the muon and the incoming neutrino is approximately $0.7^{\circ} \times (E_{\nu} / 1 \; \mathrm{TeV})^{-0.7}$~\cite{doi:10.1146/annurev.nucl.50.1.679}, which is typically smaller than the uncertainty introduced by our incomplete knowledge of light propagation in the natural detection medium. Therefore, both are considered colinear. The muon directional-reconstruction capabilities of current detectors have been validated through the study of the cosmic-ray ``Moon shadow''~\cite{Aartsen:2013jdh, Albert:2018yoj}, which delivers a typical angular resolution $\lesssim 1^{\circ}$. For comparison, events resulting from NC interactions and CC interactions of $\nu_{e}$ and $\nu_{\tau}$ have angular resolutions of $5^{\circ}$ - $15^{\circ}$~\cite{Aartsen:2019epb, 2017showerrecoAntares}. These events, called ``cascades,'' are neutrino-induced showers that are $<10$~m, smaller than the string spacing of the detector, and combined with short scattering lengths of light, they look like cascades of photons emitted from a single point in the detector.

The search for astrophysical neutrinos is conducted in a background-dominated regime, where the main backgrounds are muons and neutrinos from cosmic-ray interactions in the atmosphere. These backgrounds have distinctive spectral, directional, and flavor characteristics that allow their separation from a putative astrophysical signal. In the sky above the horizon (zenith angle $\theta < 90^{\circ}$), down-going muons represent the main background, although their soft spectrum above a few hundred GeV (asymptotically approaching $\propto E_\nu^{-3.7}$) limits their energies to $\lesssim 100$ TeV. Atmospheric neutrinos can be detected over $4\pi$ steradians, with ``conventional'', mostly muon, neutrinos being produced in pion and kaon decays with a soft $\propto E_\nu^{-3.7}$ spectrum~\cite{2016crpp.book.....G}. The much subdominant decay of charmed $D^{\pm}$ mesons introduces a predicted~\cite{Enberg:2008te}, but yet undetected, ``prompt'' neutrino background with a harder spectrum $\propto E_\nu^{-2.7}$ at $E_{\nu} \sim 100$ TeV and equal electron and muon neutrino contributions. By contrast, an astrophysical neutrino flux is expected to be in near equipartition of neutrino flavors, as they oscillate during propagation over very long distances~\cite{Learned:1994wg}. 

%% file: diffuse.tex
\section{DIFFUSE EMISSION}\label{sec:diffuse} %(4-5 pages total)
\subsection{Diffuse Neutrino Observations}
\begin{figure}[h]
\includegraphics[width=4in]{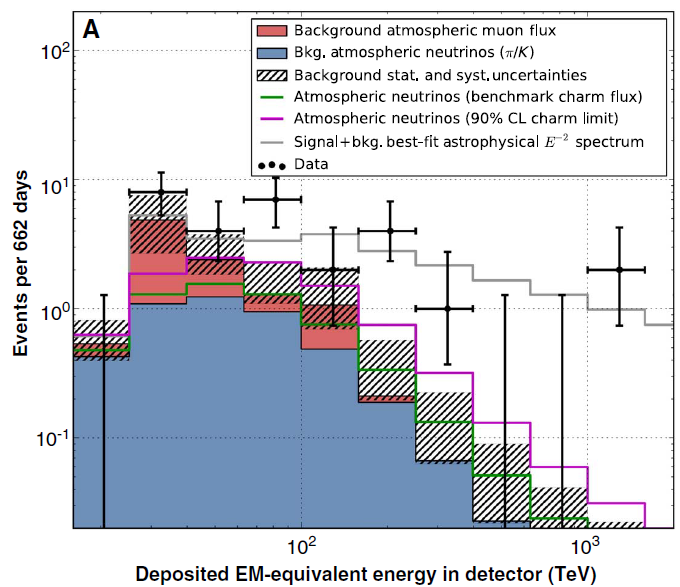}
\caption{Distributions of the deposited energies of the observed events in the IceCube detector compared to model predictions, from the 2013 discovery paper~\cite{Aartsen:2013jdh}.}
\label{HESE}
\end{figure}

The most significant breakthrough in the last decade for extragalactic neutrinos is their actual observation. While they had been long hypothesized, the first observation of celestial emission of neutrinos in the TeV to PeV range occurred in 2013. The identification of PeV-energy neutrinos by IceCube~\cite{Aartsen:2013bka}, which were incompatible with the soft atmospheric neutrino background with $2.8\sigma$, led to the development of an analysis targeting high-energy events with their interaction vertex contained within the IceCube instrumented volume. By defining a veto region in the outer parts of the detector, atmospheric muon and neutrino backgrounds were suppressed to a handful per year~\cite{PhysRevD.90.023009}, uncovering the astrophysical spectrum at the highest energy tail. While muon neutrinos interacting outside the detector have the advantage of higher statistical power, they were selected against in the discovery analysis~\cite{Aartsen:2013jdh} due to the high background expected in that sample. The resulting data set is dominated by cascade events that have superior energy resolution at 10-20\% as the light deposition is contained in the detector. Despite the limited angular resolution of the sample, a search for point sources was performed which yielded no statistically-significant evidence of directional clustering. 
In fact, the all-sky nature of the arrival direction along with a hard energy spectrum (best fit flux was $\propto E_\nu^{-2.2}$) indicates that extragalactic neutrino flux dominates this diffuse emission.

%IceCube observed excess neutrinos in their energy spectrum;  a high energy tail that was unable to be explained by backgrounds. The neutrinos selected for this analysis are not those that create muons in $\nu_{\mu}+\bar{\nu}_{\mu}$ CC interactions outside the detector, as previously described. While analyses that localize sources in the sky uses such muons for their superior angular resolution and statistical power, such a data set is background dominated. 
%For this discovery analysis, a higher purity data set was created by selecting very high energy events with light emission starting in the inner parts of the detector. 

%Most of the events selected here are what are known as "shower" type events from electron and tau neutrinos interacting via charge-current, or neutral-current interactions of all neutrino types. They result in electromagnetic showers of tens of meters, much below the spacing of most PMTs in the detector. Combined with a low scattering length in the South Pole glacier ice, they look like cascades of photons emitted from a single localized point in the detector. While the direction the neutrino originated from in the sky are harder to reconstruct, they are superior in energy resolution at 10-20\% when the shower is totally contained in the detector as they are here. The angular resolution of these events are $5-15^{o}$, but an analysis was performed to show there was no statistically significant directional clustering. 

\begin{table}[h]
\tabcolsep7.5pt
\caption{Diffuse Flux Summary}
\label{tab1}
\begin{center}
\begin{tabular}{l|l|c}
\hline
Analysis  &$\gamma$ &$\Phi_{0}$ at 100~TeV\\
\hline
Discovery updated~\cite{PhysRevD.104.022002}      &$2.87\pm0.20$ &$6.37^{+1.47}_{-1.62}$ \\
Upgoing tracks~\cite{abbasi2021improved} &$2.37\pm0.09$ &$4.32^{+0.75}_{-0.78}$ \\
Showers~\cite{PhysRevLett.125.121104} &$2.53\pm{0.07}$ &$4.98^{+0.75}_{-0.81}$ \\
\hline
\end{tabular}
\end{center}
\begin{tabnote}
Best-fit single power-law flux parameters of various diffuse flux analysis channels. Values are for an all-flavor combined flux assuming 1:1:1 flavor ratio, neutrino and anti-neutrino combined, where $\Phi_{0}$ is the flux normalization at 100 TeV in units of $10^{-18}$ GeV$^{-1}$ cm$^{-2}$ s$^{-1}$ sr$^{-1}$. 
\end{tabnote}
\end{table}
%NOT INCLUDED
%MESE 2 yr
%inelasticity https://arxiv.org/abs/1808.07629
%flavor ratio https://arxiv.org/abs/1502.03376

%In fact, the all-sky nature of the arrival direction of these neutrinos along with a hard energy spectrum, best fit was $\propto E^{-2.2}$, indicate that there must be at least some extragalactic component of this diffuse emission. 
The statistical significance of the initial discovery (for which the neutrino spectrum is shown in Figure \ref{HESE}) was 4.8$\sigma$; equivalent to a chance probability of $2\times10^{-6}$ for background fluctuations causing such observations.  Since then, the significance of the detection has increased as IceCube accumulates statistics, and the signal has also been observed in multiple analysis channels. The diffuse astrophysical neutrino flux is typically characterized using a power-law of the form $\Phi_\nu (E_\nu) = \Phi_{0} (E_\nu/E_{0})^{-\gamma}$ where the parameters are: the spectral index $\gamma$ and the flux normalization $\Phi_{0}$ quoted at the normalization neutrino energy $E_{0}$. The flux can be quoted as per-flavor or all flavors combined assuming a 1:1:1 flux ratio among the three flavors (with $\nu+\bar{\nu}$ combined), and usually normalized at $E_{0}=100$~TeV. 
Fermi shock acceleration of cosmic rays typically predicts the flux of neutrinos to follow a power law with $\gamma = 2$~\cite{Drury:1983zz}. 
Most astrophysical objects are, however expected to have more complex neutrino emission profiles with an energy dependent spectral index. Nonetheless, with statistical limitations of observations, and because an all-sky diffuse flux, which undoubtedly is the sum of many astrophysical source emissions, is being characterized as one flux, a single power-law fit remains the first test in characterising the diffuse astrophysical neutrino flux. 

The discovery analysis used 2 years of IceCube data. Since then, this analysis channel has been updated a few times with the most recent analysis using 7.5 years of data~\cite{PhysRevD.104.022002}. Another approach is to select for all tracks, not just ones originating within the detector. The track events analysis uses 9.5 years of IceCube events exclusively from the Northern sky, thus suppressing astrophysical muons, and most effectively probes energy scales of hundreds of TeV. 
Finally, using shower events exclusively suppresses astrophysical muons almost completely thus being able to push the target energies lower. This analysis uses 6 years of all-sky IceCube events and most effectively probes tens of TeV. These results are summarized in \textbf {Table \ref{tab1}}. Differences between these observations can be due to many factors. 
The astrophysical diffuse flux may have different characteristics at different energy ranges, and the four analyses probe different source components. 
The flux is likely to vary depending on the part of the sky observed, and the muon neutrino analysis only uses events from the northern sky, where as the others see the whole sky. 
The flux characterization may also depend on the neutrino flavor. Different neutrino production mechanisms at astrophysical sources lead to different neutrino flavor compositions. Neutrino  oscillations over cosmological distances will push towards an even flavor ratio but possibly not completely.

\subsection{Multimessenger Connection}
One of the most important conclusions from the IceCube observation of high-energy neutrinos is that the diffuse neutrino flux in the 0.1-1~PeV range is comparable to those of the diffuse gamma-ray background flux in the sub-TeV range~\cite{Fermi-LAT:2014ryh} and the ultra-high-energy cosmic-ray (UHECR, $E \gtrsim 10^{18.5}$ eV) flux.
This implies that the energy generation rate densities of high-energy neutrinos, high-energy gamma rays, and UHECRs are all comparable to $\sim{10}^{44}-10^{45}~{\rm erg}~{\rm Mpc}^{-3}~{\rm yr}^{-1}$~\cite{Murase:2018utn}. This gives profound constraints on the candidate sources of high-energy neutrinos, and may indicate the possible connection among three cosmic particle channels~\cite{Murase:2016gly,Yoshida:2020div}. 
For ``grand-unified scenarios'', which aim to simultaneously explain the diffuse fluxes of all three messengers, cosmic-ray reservoirs are among the most promising candidate source classes~\cite{Murase:2013rfa}. 

%\begin{figure}[h]
%\includegraphics[width=\textwidth]{figs/spectra.pdf}
%\caption{The multi-messenger context of the diffuse neutrino spectrum: the IceCube all-sky neutrino flux~\cite{IceCube:2017zho} is compared to the isotropic gamma-ray flux observed 
%by \emph{Fermi}-LAT~\cite{Fermi-LAT:2014ryh} and the flux of ultra-high-energy cosmic rays reported by the Pierre Auger and Telescope Array observatories~\cite{Deligny:2020gzq}.}
%\label{f:spectra3}
%\end{figure}

\begin{figure}[h]
\includegraphics[width=\textwidth]{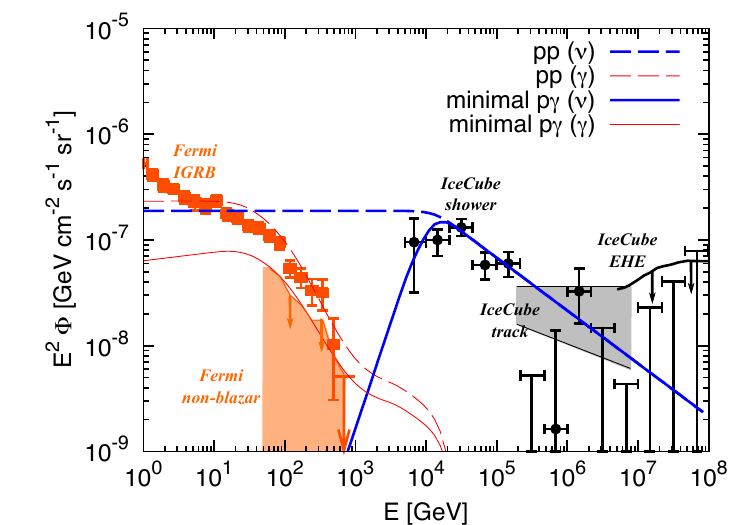}
\caption{Neutrino and gamma-ray connection of all-sky neutrino~\cite{PhysRevLett.125.121104,IceCube:2016umi, IceCube:2018fhm} and gamma-ray fluxes~\cite{Fermi-LAT:2014ryh}. For generic models to explain the IceCube data in the 10-100~TeV range, whether neutrinos are produced via $pp$ or $p\gamma$ process, gamma rays associated with neutrinos violate the non-blazar component of the extragalactic gamma-ray background measured by {\it Fermi}. Adapted from~\cite{Murase:2015xka}.}
\label{f:spectra}
\end{figure}

The neutrino and gamma-ray connection (\textbf{Figure~\ref{f:spectra}}) is unavoidable because neutrinos should be accompanied by hadronic gamma rays. 
Murase et al.~\cite{Murase:2013rfa} confronted the IceCube data above 0.1~PeV with the isotropic diffuse gamma-ray background measured by {\it Fermi}, and placed constraints on the dominant sources of high-energy cosmic neutrinos. In particular, if the neutrinos are produced by $pp$ interactions and the sources are transparent to gamma rays, the spectral index is constrained to be $\gamma\lesssim2.1-2.2$. This also implies that a steep spectrum of the diffuse neutrino flux cannot be readily reconciled with the diffuse gamma-ray flux seen by {\it Fermi}. 
Indeed, the latest shower data in the 10-100~TeV have deepened the mystery about the origin of high-energy cosmic neutrinos~\cite{PhysRevLett.125.121104}. Detailed multi-messenger analyses indicate that the hadronic gamma-ray flux associated with the diffuse neutrino flux in the 10-100~TeV range violate the non-blazar (non-point-source) contribution of the diffuse gamma-ray flux~\cite{TheFermi-LAT:2015ykq}. This suggests the existence of a class of high-energy neutrino sources that are opaque for GeV-TeV gamma rays, which can be naturally realized if neutrinos are produced by $p\gamma$ interactions~\cite{Murase:2015xka,Capanema:2020rjj,Capanema:2020oet}.  

%% file: xgal_exp.tex
\section{EXTRAGALACTIC SOURCES: EXPERIMENTAL RESULTS}\label{sec:xgal_results}

In this section, we highlight significant experimental results of the last decade in searching for extragalactic sources of neutrino emission.

Searches for correlation between neutrino data and known sources can take place by analyzing source positions one at a time, or by stacking sources belonging to the same catalog class to look for correlations simultaneously. The first method provides clear answers on whether individual sources are responsible for high-energy neutrino emission, but statistical trials must be taken into account for the look-elsewhere effect which can weaken statistical significances of observations if many sources are considered. Each collaboration keeps a list of potential neutrino sources~\cite{Aartsen:2019fau, Albert:2017ohr, Aartsen:2020xpf} with individual selection criteria based on electromagnetic observations, mostly from \emph{Fermi}-LAT and TeV source catalogs (\emph{e.g.,} 4FGL~\cite{Fermi-LAT:2019yla}, 3LAC~\cite{Fermi-LAT:2015bdd}, TeVCat~\cite{Wakely:2007qpa}, etc.). Active galactic nuclei (AGN) belonging to different subclasses such as BL Lac objects and flat spectrum radio quasars (FRSQs) dominate this list. The stacking method looks at a class of sources together, and therefore would be the first to see a signal if several weak sources emerge. However, this type of analysis is sensitive to the relative weighting of the fluxes assumed in the search. For example, weighing all sources in a catalog equally provides very different outcomes in most stacking analyses compared to, say, weighing them by their gamma-ray flux (\emph{i.e.,} assuming that the neutrino flux of each source in the catalog scales proportionally to its observed gamma-ray flux). Thus stacking analyses only answer hypotheses of not only a particular class of sources being neutrino emitters, but also the specific relative neutrino fluxes emitted.

Another way to divide these analyses is between time-dependent and time-integrated searches. Some searches, such as searches of neutrinos coincident with gamma-ray bursts (GRBs), naturally have a time dependence in the analysis. An external event triggers a search window in time. Other sources that are known to continuously emit gamma rays would be searched in neutrinos in a time-integrated way aiming to accumulate enough data so that the neutrino signal surpasses the background expectations. Much like the time-integrated analyses, these bursts or flares can be searched individually, taking into account trials, or stacked, as described in the paragraph above.

Finally, an inherently multimessenger approach is to issue neutrino alerts of significant events~\cite{2012AlertAN,2017realtimeIC} or receive alerts from other observatories to follow up in neutrinos in realtime~\cite{Dornic:2019ag,Suvorova:2021ou,2021FARIC}. In this approach, as will be described in Section~\ref{sec:mma_picture}, neutrino signals are correlated in space and time with photons or gravitational waves, which improves the sensitivity to joint emitters of these messengers.
Additionally full-sky scans of neutrino hotspots are performed to search for sources without any assumptions of counterparts. This can be done by time-integrated hotspot searches~\cite{Albert:2017ohr,Aartsen:2018ywr,Aartsen:2019fau, Aartsen:2019epb,Aartsen:2020xpf} and untriggered flare searches~\cite{2021allskyflare}.

Correlation searches between neutrino observations and known extragalactic sources have been performed for several decades~\cite{Spiering:2019}. However, to date, only two sources have emerged as tantalizing neutrino sources: the active galaxies TXS 0506+056 and NGC 1068.

\subsection{Blazars}
The IceCube Neutrino observatory issues realtime alerts to the astronomical community when high-energy neutrino events of likely astrophysical origin are recorded in the detector~\cite{Blaufuss:2019fgv}. 
On September 22, 2017, such an alert~\cite{kopper2017grb} was issued due to the detection of a neutrino event, IceCube-170922A. The position of the event was found to be coincident in time and direction with the gamma-ray blazar TXS 0506+056, as shown in \textbf{Figure \ref{f:spectra2} (Left)}, which at the time was flaring in gamma-rays and X-rays. Further investigation of IceCube archival data from this location found evidence of a neutrino ``flare'' from September 2014 to March 2015. 
The coincidence of such a neutrino alert event arriving from a flaring source in \emph{Fermi}-LAT constituted a 3$\sigma$ evidence for neutrino emission from the direction of TXS 0506+056~\cite{IceCube:2018dnn}. 
Independently, the neutrino flare of 2014/15 represents a 3.5$\sigma$ observation~\cite{doi:10.1126/science.aat2890}. This was the first compelling evidence of any high-energy neutrino source in the history of multimessenger astronomy.
 
Blazars have been studied using a stacking approach~\cite{2015blazarANT,2017blazarIC,Hooper:2018wyk,Smith:2020oac,2021stackingANT} but correlations have not been observed, even when TXS 0506+056 is present in the catalog since it becomes one of many sources considered simultaneously. When neutrino flares are searched over the entire sky~\cite{2021allskyflare}, the 2014/2015 flare becomes statistically insignificant due to the look-elsewhere effect. This highlights the advantage neutrino source searches gain by having an external trigger from another messenger that singles out a source of interest.

\subsection{Other Types of Active Galactic Nuclei}
By 2018, IceCube observed an excess in their source list search at the location of the Seyfert II galaxy NGC 1068~\cite{Aartsen:2019fau} at the 2.9$\sigma$ level. 
This source was also within the extended region of the most significant hotspot in the Northern hemisphere in the full-sky scan, as shown in \textbf{Figure \ref{f:spectra} (Left)}. This was the first time a time-integrated search had resulted in a hotspot at this level of statistical significance. 
NGC 1068, at a $\sim14$ Mpc distance, is the most luminous Seyfert II galaxy detected by \emph{Fermi}-LAT, making it a not surprising source to emerge this way. The analysis used 10 years of IceCube data, illustrating that cumulative long-term neutrino data can deliver evidence of sources, in addition to time-domain search techniques such as those used to identify TXS 0506+056. As more data accumulates, stronger evidences of more sources are likely to emerge in the future.

Perhaps comfortingly, another location that emerges as a hotspot in the IceCube full-sky scan of 10 years of time-integrated data is TXS 0506+056. The data used here encompasses the alert event IceCube-170922A and the 2014/2015 flare period. A comparison of how the time-integrated significance grows compared to time-dependent behavior of neutrinos from sources will become increasingly interesting for many sources.

\subsection{Gamma Ray Bursts and Other Transients}
Before neutrino telescopes were operational at the large scales they are today, the most favorable sources were considered to be GRBs. However, by 2012, IceCube established an absence of energetic neutrinos associated with GRBs~\cite{2012Nature}, but the limits were still consistent with the standard theoretical predictions~\cite{Li:2011ah,Hummer:2011ms,He:2012tq}.
%at at least a factor of 3.7 below the historically standard flux predictions~\cite{2012Nature}. 
GRBs are stacked in time and position or analyzed individually to search for coincident neutrinos, where neutrinos are analyzed to come on-time or with a time shift. The well-localized nature of GRBs in time and position give neutrino analyses a nearly background-free opportunity to search for coincident neutrinos. Despite this highly favorable search condition, and continuous improvements in analysis sensitivities, no correlations have been found~\cite{2013ANTARESGRB,2015ICGRB,Aartsen:2016qcr,Adrian-Martinez:2016xij,Aartsen:2017wea,2017ANTARESGRB,2020ANTARESGRB}, which may point to more conservative predictions or other models predicting lower-energy neutrino emission~\cite{Murase:2019tjj}.  
%making this one of the biggest surprises of the last decade.
%ignoring this https://arxiv.org/abs/2011.11411

Supernovae, which were established as MeV neutrino emitters, are also promising sources of high-energy neutrinos. 
Not only follow-up searches~\cite{IceCube:2015jsn} but also stacking searches~\cite{Senno:2017vtd,Esmaili:2018wnv,IceCube:2021oiv} have been performed for broadline Type Ibc and Type IIn supernovae, but no significant excess has been found.   

Tidal disruption events (TDEs), where a star is ripped apart by tidal forces in the vicinity of a supermassive black hole, have recently been suggested to correlate with neutrino emission~\cite{2021TDE} due to a neutrino alert event in the vicinity of a very bright TDE. Previous stacking analysis of multiple TDEs from IceCube~\cite{2019TDEIC}, however, did not observe correlation of statistical significance, nor a time-integrated search of TDEs from ANTARES~\cite{2021TDEANT}. 
%These observations are not in contradiction under some models since different assumptions are made in each analysis.
%I am pretty uncomfortable including this paragraph here about TDEs

\subsection{Sources of Other Messengers}
Gravitational wave alerts are followed up by neutrino telescopes to search for coincident neutrinos. The famous multimessenger detection of the gravitational-wave (GW) event GW170817 associated with a GRB and a kilonova/macronova was followed up by neutrino telescopes~\cite{GBM:2017lvd, ANTARES:2017bia, Avrorin:2018fzl}. Other GW events -- GW170104\cite{Albert:2017obm}, GW15226~\cite{ ANTARES:2017iky} and GW150914~\cite{Adrian-Martinez:2016xgn} -- were also analyzed in search for neutrino counterparts, as well as entire runs of LIGO and Virgo~\cite{Albert:2018jnn,2020ICGW, abbasi2021probing, 2020ANTARESGW}. No associated neutrino emission was identified in these searches, and upper limits on the neutrino luminosity of these events were derived. 

UHECRs can also be correlated with neutrinos under the assumption that their arrival directions are not completely scattered during propagation by Galactic and intergalactic magnetic fields. As UHECRs interact with ambient photons and matter at their acceleration sites or during propagation, the neutrino emission may trace the positions of their sources which could be revealed in a joint analysis~\cite{2013ANTARESUHECR,Aartsen:2016ngq}. No correlation between UHECRs and neutrinos has been identified so far.

As both GW and UHECR correlation studies have yielded null results, it remains that photons are the only other messenger where evidence for a neutrino correlation is observed.

%% file: xgal_srcs.tex
\section{EXTRAGALACTIC SOURCES: CANDIDATES}
\subsection{Active Galactic Nuclei}
AGNs are among the most promising candidate sources of high-energy cosmic rays and neutrinos. They typically host supermasssive black holes with masses of $\sim10^6-10^9~M_\odot$, and the accretion onto the black hole powers radiation from the accretion disk and its coronae as well as winds. The black holes are believed to rapidly spin, which also drives powerful relativistic jets.  

\subsubsection{Blazars and TXS 0506+056}
AGN with powerful jets may point to the Earth, which are called blazars.  
Their emission is strongly beamed and non-thermal, and the spectral energy distributions are composed of two humps. The low-energy hump is attributed to synchrotron emission from electrons accelerated in the jet, while the origin of the high-energy hump has been under debate. In the leptonic scenario, gamma rays are explained as inverse-Compton emission. In the hadronic scenario, cosmic-ray-induced cascade emission~\cite{Mannheim:1995mm} and/or cosmic-ray synchrotron emission~\cite{Aharonian:2000pv,Mucke:2000rn} are responsible for the observed gamma rays. In either scenario, not only electrons but also ions are accelerated, and lepto-hadronic models have been widely investigated. 

Accelerated ions produce neutrinos through $pp$ and $p\gamma$ interactions, and the latter process is typically more important in blazars. 
For BL Lac objects, target photons are mainly synchrotron photons from relativistic electrons that are co-accelerated together with ions. External radiation fields, which can be broadline emission in the UV range, dust emission at the IR band, and scattered disk-corona emission in the UV and X-ray range , play dominant roles in neutrino emission from FSRQs. The neutrino spectrum is predicted to be hard if the cosmic-ray spectrum follows $\propto E^{-2}$ or somewhat steeper. Blazars are typically EeV neutrino emitters if these are the sources of UHECRs (see a review, e.g., Ref.~\cite{Murase:2015ndr}), and the limits on extremely-high-energy ($E_{\nu} \gtrsim 10$ PeV) neutrinos~\cite{IceCube:2018fhm} have placed important constraints on the models.  

\begin{figure}[h]
\includegraphics[width=1.0\textwidth]{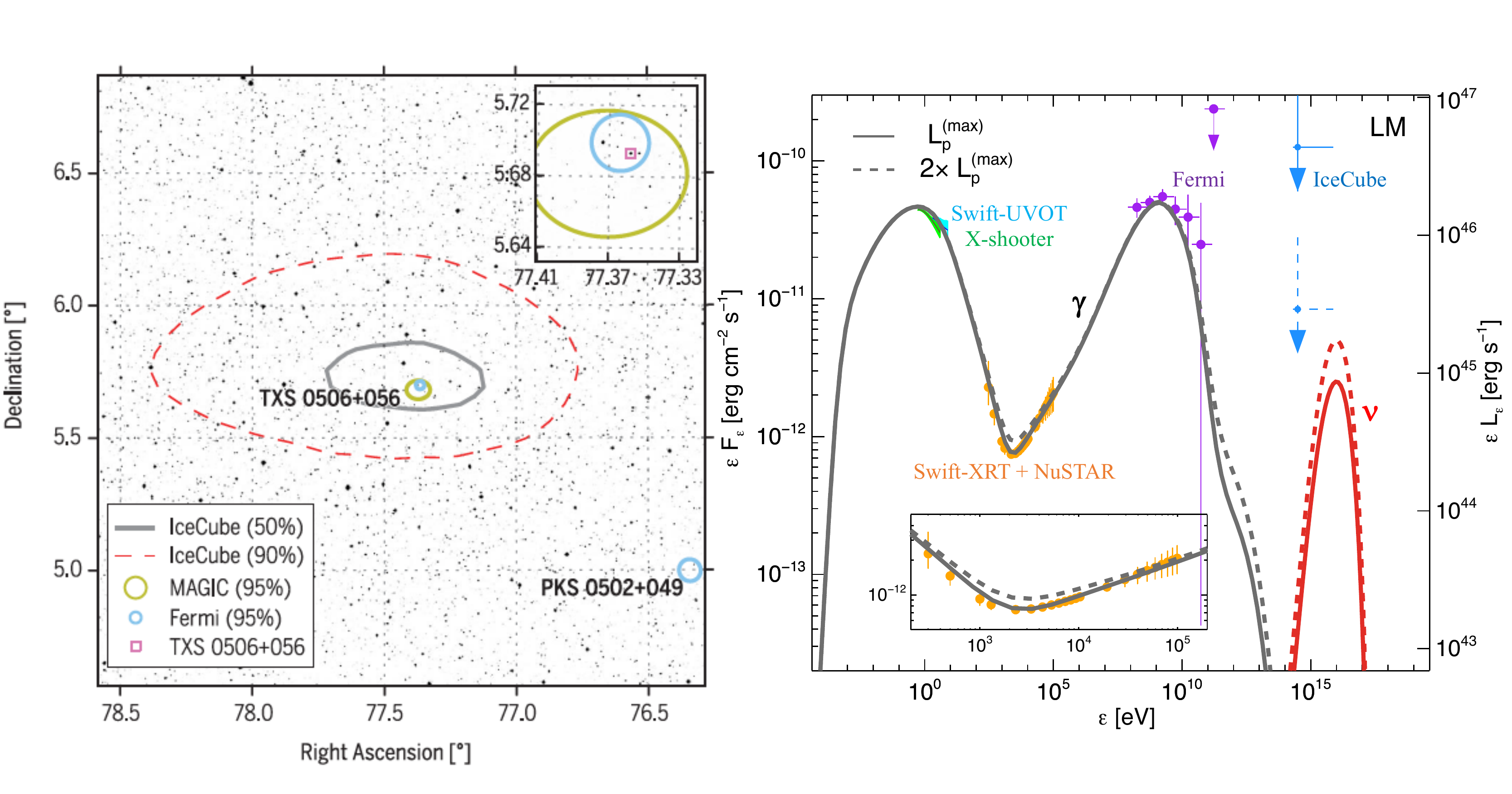}
\caption{(Left) Positions of IceCube-170922A with angular uncertainties and TXS 0506+056~\cite{doi:10.1126/science.aat2890}. (Right) Multimessenger spectra of TXS 0506+056. All-flavor neutrino flux upper limits are shown for 0.5 (solid blue line) and 7.5 (dashed blue line) years, respectively (adapted from~\cite{Keivani:2018rnh}).}
\label{f:spectra2}
\end{figure}

TXS 0506+056 is a blazar with an intermediate luminosity, which may be classified between a BL Lac and a FSRQ, although it may be a masquerade blazar~\cite{Padovani:2019xcv}.
Detailed numerical modeling based on a single emission zone model suggests that it is challenging to build a concordance picture of multimessenger emission from the optical, X-ray, gamma-ray bands to in neutrinos. For the 2017 multimessenger flare, the \emph{Swift} and \emph{NuSTAR} X-ray data clearly show a valley in the spectrum, by which the neutrino and cosmic-ray luminosities are strongly constrained (see \textbf{Figure~\ref{f:spectra2} right}). 
This is because gamma rays, electrons and positrons associated with neutrinos inevitably initiate electromagnetic cascades with a broad spectrum over the wide energy range, which would fill the valley if the neutrino flux was as high as the upper limit~\cite{Keivani:2018rnh,MAGIC:2018sak,Gao:2018mnu,Cerruti:2018tmc,Gasparyan:2021oad}.   
The interpretation is more challenging for the 2014-2015 neutrino flare, which did not show the coincident flare in either gamma rays or X-rays and the electromagnetic cascade flux associated with the neutrino flux violates the X-ray and/or gamma-ray data~\cite{Murase:2018iyl,Rodrigues:2018tku,Reimer:2018vvw,Petropoulou:2019zqp}. 
Alternative models, which include the neutral beam model and two-zone models with the gamma-ray hidden region, have been proposed to avoid the constraints from electromagnetic cascades~\cite{Murase:2018iyl,Zhang:2019htg,Xue:2019txw}. 
However, the situation is still controversial. Although a few more coincidences have been reported~\cite{Kadler:2016ygj,Giommi:2020viy,Petropoulou:2020pqh,Rodrigues:2020fbu,Oikonomou:2021akf}, further observations and theoretical investigations are necessary to establish blazars as the sources of high-energy neutrinos. 

\subsubsection{Non-Jetted AGN and NGC 1068}
Most of AGN are radio quiet without possessing powerful jets (although they may have weak jets that are often inferred by radio observations), and they are typically Seyfert galaxies and quasars. The vicinity of black holes has been discussed as possible sites of particle acceleration and resulting neutrino emission~\cite{Ber77,Eichler:1979yy}. Historically, accretion shocks have been considered as possible particle acceleration sites~\cite{PK83}, and the diffuse neutrino flux in this scenario has been calculated~\cite{Stecker:1991vm}. However, the existence of accretion shocks is in question, and the model has been constrained by neutrino observations. 

In the standard disk-corona picture, X-ray emission is attributed to Comptonized disk photons, where the high-temperature corona is expected to form around the disk through magnetic dissipation. The coronal region is magnetized and turbulent, where ion acceleration can occur through magnetic reconnections and stochastic acceleration. For magnetically-powered coronae in Seyfert galaxies and quasars, high-energy neutrinos in the 10-100~TeV range are predicted, which can account for the all-sky neutrino flux in this medium-energy range~\cite{Murase:2019vdl}. The Bethe-Heitler pair production is important as an energy loss process of cosmic rays, and MeV gamma rays are unavoidably generated through electromagnetic cascades.  

\begin{figure}[h]
\includegraphics[width=1.0\textwidth]{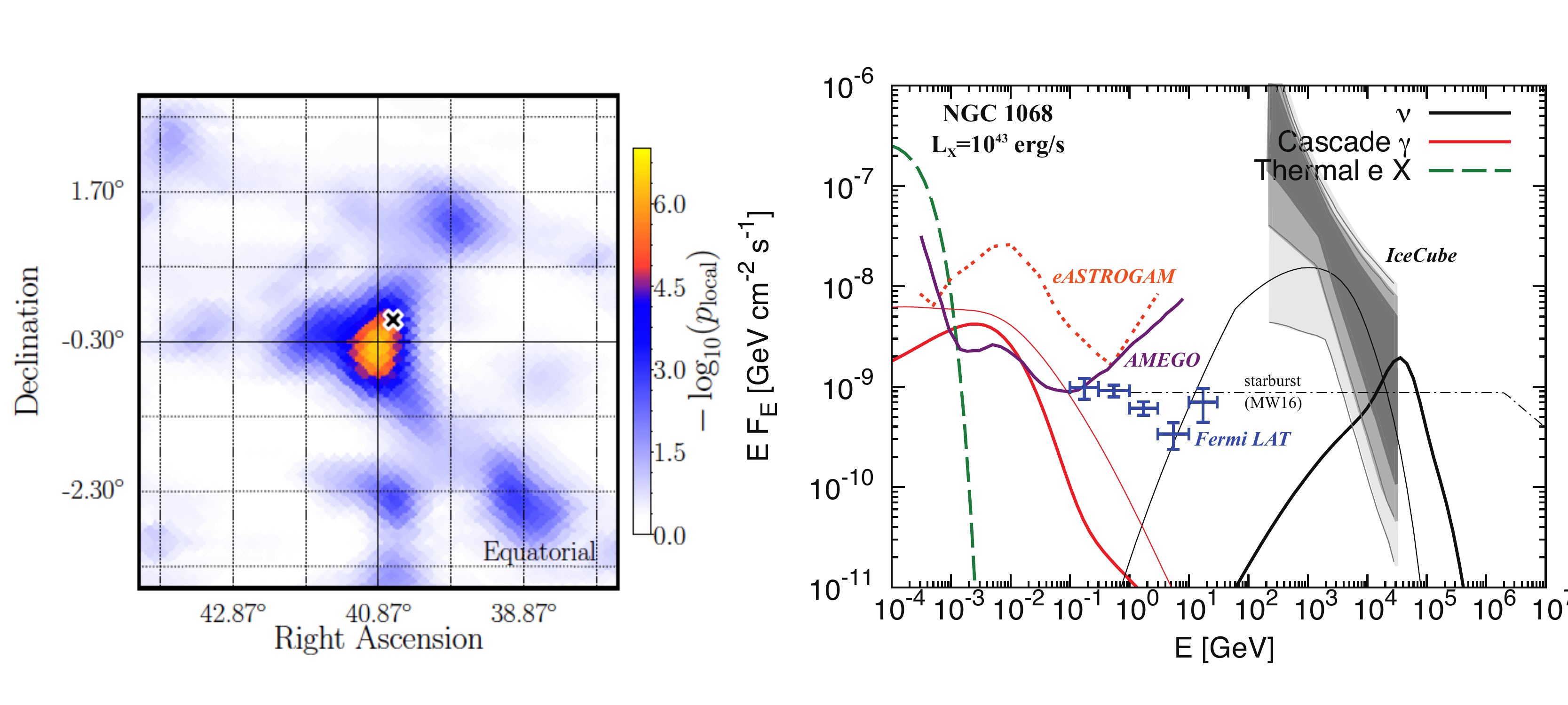}
\caption{(Left) The significance map of high-energy neutrinos, where the position of NGC 1068 is indicated by the cross~\cite{Aartsen:2019fau}. (Right) Multimesssenger spectra of NGC 1068~\cite{Murase:2019vdl}, where thin (thick) curves for neutrinos and cascade gamma rays represent models optimized for NGC 1068 (all-sky diffuse neutrino intensity in the 10-100~TeV range).}
\label{f:spectra5}
\end{figure}

The association of high-energy neutrinos with NGC 1068 is intriguing in several aspects (\textbf{Figure~\ref{f:spectra5} right}). First, NGC 1068 is a starburst coexisting with an AGN. Second, NGC 1068 is a Compton-thick Seyfert galaxy, in which X-rays are largely absorbed. 
The IceCube observation of NGC 1068 is consistent with either accretion shock or magnetically-powered corona model~\cite{Murase:2019vdl,Inoue:2019yfs,Anchordoqui:2021vms}. 
In the disk-corona model, NGC 1068 is predicted to be the brightest neutrino source in the northern sky and the model can critically be tested by IceCube and IceCube-Gen2, and a few brighter sources in the southern sky are expected to be observed by KM3Net and Baikal-GVD~\cite{Kheirandish:2021wkm}. 
Multimessenger observations are also important, and in particular MeV gamma-ray observations will serve as a complementary probe of particle acceleration in the vicinity of supermassive black holes. 

\subsection{Violent Transients}
\subsubsection{Tidal Disruption Events}
As a star is torn apart by a supermassive black causing a TDE, about half ot its mass is unbound, while the other half is expected to accrete onto a black hole or go outward as winds. TDE thermal emission, which can be produced by accretion disks, interactions of tidal streams, and winds (through reprocessing), has been observed in the optical and X-ray bands. Some of the TDEs are believed to have powerful jets, with Sw 1644+57 being an example of a powerful TDE with strong X-ray emission~\cite{2011Sci...333..203B}.  
Since then, jetted TDEs have been actively discussed as the sources of high-energy neutrinos~\cite{Wang:2015mmh,Senno:2016bso,Dai:2016gtz,Lunardini:2016xwi}. 
Non-jetted TDE neutrino emission has also been considered, where neutrinos may come from accretion disks, coronae, and sub-relativistic outflows interacting with the TDE debris or clouds~\cite{Murase:2020lnu,Hay21}.

AT 2019dsg was a luminous TDE, which was associated with IceCube-191001A~\cite{2021TDE}. The optical luminosity at the peak is $L_{\rm OUV}\sim10^{44}~{\rm erg}~{\rm s}^{-1}$, and the neutrino event was observed 150~days after the peak. This TDE was detected in the radio and X-ray bands, and accompanied by infrared echo emission. 
AT 2019fdr was another very luminous optical transient, associated with IceCube-200530A~\cite{Reusch:2021ztx}. The optical luminosity at the peak is $L_{\rm OUV}\sim10^{45}~{\rm erg}~{\rm s}^{-1}$, and the neutrino was seen 320~days after the peak. This TDE was also accompanied by a radio counterpart. These detections may hint at TDEs as one of the contributors to the all-sky neutrino flux.

\subsubsection{Long Gamma-Ray Bursts and Supernovae}
A massive star with a mass of $\gtrsim8~M_\odot$ ends its life with a violent explosion, which is a supernova. 
It is accompanied by emission of thermal neutrinos in the 10~MeV range, and a fraction of the released gravitational binding energy is transferred to the kinetic energy, which is typically $\sim10^{51}$~erg. The ejecta typically expands with a high velocity of $\sim(3000-10000)~{\rm km}~{\rm s}^{-1}$, forming a strong shock, where the diffusive shock acceleration mechanism operates. Supernova remnants are believed to be primary candidate sources of Galactic cosmic rays, and they have been observed in gamma rays.  

Supernovae are natural sources of optical photons that originate from radioactive nuclei, shock heating, and perhaps energy injection by the central engine. Recent optical surveys revealed that it is common that supernova progenitors are accompanied by the massive eruption of circumstellar material (CSM) or inflation of stellar envelope. It has been suggested that such a system is a promising source of high-energy neutrinos~\cite{Murase:2010cu,Katz:2011zx,Murase:2017pfe}. For ordinary Type II supernovae, IceCube, KM3NeT and Baikal-GVD may detect $\sim100-1000$ neutrinos in the TeV range, and even millions of sub-TeV neutrinos can be detected if Betelgeuse explodes~\cite{Murase:2017pfe}. Type IIn supernovae are powered by collisions with their circumstellar medium, and neutrinos from nearby supernovae are detectable~\cite{Murase:2010cu,Petropoulou:2017ymv}. Detecting high-energy neutrinos is important to reveal ion acceleration in the early stage of supernovae and reveal the origin of Pevatrons.   

GRBs are among the brightest explosive phenomena in the universe.
For long GRBs, their prompt emission lasts for $\sim10-1000$~s, which is believed to originate from relativistic jets that are launched by a black hole or magnetar. Ions can be accelerated by jets, and high-energy neutrinos can be produced mainly via $p\gamma$ interactions~\cite{Waxman:1997ti}. The predicted fluxes have large uncertainties~\cite{Murase:2005hy,Hummer:2011ms,He:2012tq}, and the current limits start to give stringent constraints on the models.
In the classical scenario, prompt gamma-ray emission is attributed to synchrotron radiation from relativistic electrons that are accelerated at internal shocks. However, recent observations and theoretical studies led to alternative scenarios such as photospheric emission and magnetic dissipation scenarios. 
In the photospheric emission scenario, GeV-TeV neutrinos may be more promising~\cite{Bahcall:2000sa,Meszaros:2000fs,Murase:2013hh}. This is because cosmic-ray acceleration via the diffusive shock acceleration mechanism is suppressed when the shocks are radiation mediated, although ions can still be accelerated through the neutron conversion.  

Ions can be accelerated up to ultrahigh energies at the external reverse shock during the afterglow phase, in which EeV neutrinos can be produced~\cite{Waxman:1999ai,Dai:2000dj,Dermer:2000yd}. A significant fraction of GRBs have X-ray and ultraviolet flares, which may also dominate neutrino emission from GRBs~\cite{Murase:2006dr,Guo:2019ljp}. 

Low-luminosity GRBs and transrelativistic supernovae are regarded as intermediate objects between GRBs and supernovae, and they can be the major sources of high-energy neutrinos and UHECRs~\cite{Murase:2006mm,Gupta:2006jm,Wang:2007ya,Murase:2008mr}. They are less luminous but more common. 
Recent studies have indicated that some supernovae are powered by jets and/or magnetar winds, which may also be promising sources of high-energy neutrinos and UHECRs~\cite{Murase:2013ffa,Senno:2015tsn,Tamborra:2015fzv,Grichener:2021xeg, Murase:2009pg,Fang:2013vla}. In particular, choked jets have been suggested as the sources of the all-sky neutrino flux in the 10-100~TeV range~\cite{Murase:2013ffa,Carpio:2020app}.

\subsubsection{Short Gamma-Ray Bursts and Compact Binary Mergers}
Binary neutron star mergers, neutron star - black hole mergers, and binary black hole mergers, which are powerful sources of gravitational waves, have been discussed as potential sources of high-energy neutrinos. 
In particular, double neutron star mergers are widely believed to be the progenitors of short GRBs, and the detection of GW170817 associated with GRB 170817A supported this hypothesis. High-energy neutrinos from short GRBs have been investigated mostly in light of the jet scenario~\cite{Kimura:2017kan,Biehl:2017qen,Ahlers:2019fwz}, and neutrino emission during the extended and plateau emission phases are likely to be dominant~\cite{Kimura:2017kan}. 
High-energy neutrinos may also be produced in choked jets~\cite{Kimura:2018vvz}, as well as by winds from the remnant black hole or magnetar~\cite{Fang:2017tla,Decoene:2019eux}. 
Black hole mergers could also be neutrino emitters if there is matter around the binaries (see a review, e.g., \cite{Murase:2019tjj}).

\subsection{Cosmic-Ray Reservoirs}
Magnetized environments, in which cosmic rays are confined, provide natural sites for the production of high-energy neutrinos and gamma rays. If cosmic-ray accelerators (e.g., supernovae and AGN) are embedded in such environments, cosmic rays escaping from the accelerators may diffuse and cause $pp$ and $p\gamma$ interactions. In extragalactic space, the most promising sources, which are regarded as cosmic-ray reservoirs, are galaxies and galaxy groups. 

\subsubsection{Galaxy Clusters and Groups}
Galaxy clusters are known to have magnetic fields with $\sim0.1-1~\mu$G strength which can confine cosmic rays over cosmological timescales. It has been suggested that galaxy clusters and groups are the sources of high-energy cosmic rays around the ``second knee'' at $\sim10^{17}$~eV, a softening in the cosmic-ray spectrum, or even UHECRs. 
There are two potential acceleration sites. The first is shocks associated with large-scale structure formation, which involve accretion shocks and merger shocks. Second, cosmic rays may be accelerated by AGN and/or supernovae that are embedded in clusters and groups. Low-energy cosmic rays are expected to lose their energies via adiabatic losses, while sufficiently high-energy cosmic rays above PeV energies can escape into the intracluster medium without significant energy losses.  
Interestingly, the IceCube data above 100~TeV are consistent with earlier theoretical predictions for neutrino emission from galaxy clusters and groups~\cite{Murase:2008yt,Kotera:2009ms}. However, the accretion shock model, where cosmic rays are produced at accretion shocks around the virial radius is already disfavored because of constraints from neutrino anisotropy and radio observations~\cite{Murase:2016gly,Zandanel:2014pva,Fang:2016amf}. On the other hand, the AGN model is still viable, where cosmic rays are supplied by AGN including radio galaxies and low-mass clusters at high redshifts make a significant contribution to the all-sky neutrino flux~\cite{Murase:2013rfa,Fang:2017zjf}.  

The fact that the energy generation densities of three messenger particles are all comparable suggests that their sources may be physically connected. Indeed, galaxy clusters can trap cosmic rays, by which high-energy neutrinos below a few PeV can be explained with a hard spectrum~\cite{Fang:2017zjf,Hussain:2021dqp}. The non-blazar component of the isotropic diffuse gamma-ray background is also explained simultaneously, in which cosmogenic gamma rays give a significant contribution. High-energy cosmic rays above 100~PeV may escape, and it has been suggested that cosmic rays around the second knee may come from AGN or galaxy clusters~\cite{Murase:2008yt}. The hard spectrum of UHECR nuclei can be reproduced through the magnetic confinement and photodisintegration during the escape from the intracluster medium.   

\subsubsection{Star-Forming Galaxies}
Star-forming galaxies showing the most active star-formation are often called starbursts galaxies, and they typically have high column densities and strong magnetic fields with $\sim0.1-1$~mG. Cosmic rays supplied by supernovae or other transients related to massive stars can be confined for millions of years, and produce neutrinos and gamma rays~\cite{Loeb:2006tw,Thompson:2006np}. 
For galaxies like the Milky Way, the cosmic-ray diffusion is so effective that the cosmic-ray spectrum on Earth, $\propto E^{-2.7}$, is steeper than the injected spectrum. In starburst galaxies, the gas density is so high that cosmic rays can lose their energies via $pp$ interactions or adiabatic cooling before they diffusively escape, where the spectrum of neutrinos and gamma rays can be as hard as the observed neutrino spectrum. For starburst galaxies to be a major contributor of IceCube neutrinos in the PeV range, cosmic rays need to be accelerated beyond 100~PeV energies. This could be achieved by hypernovae or other energetic transients, stronger magnetic fields, galactic winds, and AGN embedded in highly star-forming galaxies~\cite{Murase:2013rfa,Tamborra:2014xia,Chang:2014hua,Senno:2015tra,Chakraborty:2015sta}. 
Recent studies have showed that star-forming galaxies give a significant contribution to the isotropic diffuse gamma-ray background~\cite{Roth:2021lvk,Blanco:2021icw}. However, the contribution to the all-sky neutrino flux is likely to be subdominant at least below 100~TeV, although they can still make a sizable contribution above 100~TeV given that the cosmic-ray spectrum is harder than $\propto E^{-2.2}$.  

%% file: mma_picture.tex
\section{EXTRAGALACTIC NEUTRINOS AND THEIR MULTIMESSENGER SYNERGIES IN THE COMING DECADE}\label{sec:mma_picture}
The coming decade holds the promise of fully enabling extragalactic neutrino astronomy thanks to significant advances in our observational capabilities. Efforts are currently underway that aim to increase the number, sensitivity, and energy reach of neutrino telescopes as well as establishing new multimessenger studies as additional electromagnetic and gravitational-wave observatories come online. 

\subsection{Improving Neutrino Observations}
As described in Section~\ref{sec:xgal_results}, detecting extragalactic neutrino sources via time-integrated analyses, such as the one that provided evidence for NGC 1068, are performed in a background-dominated regime although with high neutrino event rates. Enough data must accumulate before such signal emerges. Associating individual neutrino events with extragalactic sources, as with realtime alerts, is also possible if the neutrino energy is sufficiently high ($\gtrapprox$ 100 TeV) as the atmospheric background becomes subdominant at those energies; such is the case of the association between TXS 0506+056 and the IceCube-170922A event. A cubic-kilometer telescope such as IceCube may only detect $\mathcal{O}$(10) events per year across the entire sky for such individual alert events. Both approaches are limited by how fast detectors can accumulate high-energy events.

%Other approaches rely on ``cascade'' events as they have a lower atmospheric background contamination and can therefore be used to search for correlations with EM sources. Given their limited angular resolution, cascade events are best suited for extended source searches (which limits their applicability to Galactic objects) or to search for correlations with short EM transients as the atmospheric background can be significantly reduced due to the short duration of these events. The main limitation for cascade events is that most of the light deposition should occur within the detector volume to reduce a contamination from stochastic energy losses from muon events which may mimic a cascade, reducing the statistics for these searches in the high-energy range. For neutrinos, the main improvements to these searches therefore involve increasing statistics at the highest energies through the construction of additional neutrino telescopes (or the expansion of existing ones), improving the angular resolution of these detectors to enable more sensitive auto-correlation searches or cross-correlations with MMA signals, and further integrating the neutrino telescopes with the MMA community to enable correlated searches for transient or flaring extragalactic sources in near-realtime. 

Other approaches rely on ``shower'' or ``cascade'' events. Their primary advantage is the absence of atmospheric muons and atmospheric muon neutrinos detected via CC interaction as backgrounds. This makes them suitable for analyses of sources above the local horizon, where using track events would include atmospheric muon backgrounds many orders of magnitude above the neutrino signal. They are also suited for extended regions of neutrino emission, given their limited angular resolution, and for correlation studies with short electromagnetic transients as the background can be significantly reduced due to the short duration of these events. The main limitation for cascade events is their angular resolution, which is limited by understanding the local optical properties of the detector media.

For neutrino astronomy, the main improvements to these searches therefore involve increasing data collection through the construction of additional neutrino telescopes (or the expansion of existing ones), improving the angular resolution of the detectors, and further integrating the neutrino telescopes with the multimessenger astronomy (MMA) community to discover more correlations in near-realtime.

The increase in high-energy event rates will be delivered by next-generation instruments such as IceCube-Gen2~\cite{IceCube-Gen2:2020qha}, which aims to increase the instrumented volume of the current-generation IceCube to $\sim8$~km$^{3}$. In 10 years of operation IceCube-Gen2 will reach a $5\sigma$ energy flux sensitivity of the order of $10^{-12}$ erg cm$^{-2}$ s$^{-1}$ in the 0.1-1 PeV range, a similar level as current VHE $\gamma$-ray observatories in the EM domain. Additional sensitivity will be provided by KM3NeT-ARCA~\cite{Adrian-Martinez:2016fdl}, Baikal-GVD~\cite{Avrorin:2019dli}, and the recently-proposed P-ONE telescope~\cite{P-ONE:2020ljt} which are expected to reach the $\sim\mathrm{km}^3$-scale in the coming decade. 
In the $E_{\nu}\gtrapprox100$ TeV energy range neutrino absorption in the Earth becomes significant, reducing the detector effective areas to neutrinos passing through the Earth (i.e. up-going in the horizontal coordinate frame of the detector). Due to the high muon background for down-going neutrinos, and absorption effects for up-going ones, the peak sensitivity for neutrino telescopes in this energy range is near the local horizon. For mid-latitude detectors like KM3NeT and GVD this means that the region of maximum instantaneous sensitivity sweeps the sky as the Earth rotates, which is relevant for transient searches, while for IceCube and IceCube-Gen2 it will remain near the celestial equator. Given their different locations and different capabilities, searches for extragalactic neutrino sources may benefit from a joint effort that combines data from all neutrino telescopes~\cite{Schumacher:2021hhm}. 

Improvements in the angular resolution of neutrino telescopes will also enable more sensitive and critical searches for extragalactic sources~\cite{Murase:2016gly}, especially given their large number and the possibility of source confusion. 
The sensitivity of point-source searches scales roughly with the inverse of the point spread function radius of the instrument, while the probability of chance correlations for single events scales with the radius squared (i.e. it is proportional to the solid angle of the uncertainty region of the neutrino event). This means that even modest improvements in angular resolution could deliver significant sensitivity enhancements that are equivalent to longer exposure times for steady sources, and also prove vital in pinpointing neutrinos coincident with rare transient events. Detailed characterization of light propagation in the very local detector media throughout the instrumented volume will be critical for neutrino telescopes. Furthermore, increasing sampling of the Cherenkov radiation emitted by each event, made possible in a detector with more PMT instrumentation, will deliver significant improvements of angular resolution to current instruments. It is expected that next-generation telescopes will be able to reach a resolution of $\sim 0.2^{\circ}$ at 1~PeV for muon tracks, and few-degree resolution for cascades, enabling more precise searches for multimessenger counterparts to astrophysical neutrinos~\cite{KM3NeT:2018wnd,IceCube-Gen2:2021pdk}.

\subsection{The Multimessenger Landscape in the Coming Decade}
\subsubsection{New Instruments, Expanded Capabilities} Understanding the physical processes at work in neutrino sources relies on the combined study of their neutrino and EM emission, both as a function of time and energy. As neutrino telescopes are wide-field instruments with nearly $4\pi$ sky coverage, these correlated studies require wide-field EM instruments, or pointed instruments that survey large swaths of the sky with high cadence. Prompt target-of-opportunity (ToO) observations, or long term monitoring programs, of potential neutrino emitters may provide a similar temporal coverage over different timescales for a large collection of sources across the sky. 

High-energy extragalactic emitters such as AGN that are potential neutrino sources display broadband emission features spanning the entire electromagnetic spectrum, from radio to VHE gamma rays. As was discussed in previous sections, electromagnetic tracers of hadronic emission may appear at different energy ranges depending on the properties of the source region so their characterization requires broadband energy coverage. These observational campaigns, such as the one following the detection of IceCube-170922A and subsequent monitoring of TXS 0506+056~\cite{IceCube:2018dnn}, are expected to continue in the coming years as more potential neutrino sources are identified, and will only be possible if existing and upcoming EM facilities are in operation simultaneously. In the very-high-energy (VHE, E $\gtrapprox 100$ GeV) gamma-ray band, the upcoming Cherenkov Telescope Array (CTA)~\cite{CTAConsortium:2013ofs, 2021arXiv210804512G} will perform follow-up observations of neutrino alerts~\cite{CTAConsortium:2017dvg, Sergijenko:2021lwn} continuing the effort of current-generation instruments like H.E.S.S., MAGIC and VERITAS~\cite{VERITAS:2021mjg} while also carrying out the most sensitive studies of extragalactic sources in the energy range between few tens of GeV up to tens of TeV. These pointed observations will be supplemented by the monitoring capabilities of wide-field instruments currently in operation in the northern hemisphere like HAWC~\cite{Abeysekara:2017mjj} and LHAASO~\cite{Aharonian:2020iou}, and the proposed SWGO in the southern hemisphere~\cite{Hinton:2021rvp} where no such capability currently exists. For the foreseeable future, \emph{Fermi}-LAT will remain as the most sensitive instrument in the GeV band with all-sky coverage. A potential situation that should be avoided is an observational gap if \emph{Fermi} stops operation before future missions, currently in their concept phase (\emph{e.g.}~\cite{APT:2021lhj}), are launched. 

As previously discussed, an important energy range for hadronic EM studies is the hard X-ray ($>10$ keV) to MeV band. For hard X-rays, \emph{NuSTAR}~\cite{NuSTAR} is the most sensitive telescope in operation while a gap exists in the MeV range, which is expected to be closed by upcoming mission in the next decade (\emph{e.g.} AMEGO~\cite{Moiseev:2017mxg}, AMEGO-X~\cite{Fleischhack:2021mhc}, COSI~\cite{Tomsick:2021wed}). For soft X-rays ($< 10$ keV), the Neil Gehrels \emph{Swift} observatory has been the main instrument for neutrino follow-up observations given its fast repointing capability~\cite{Evans:2015qia}, while MAXI/GSC~\cite{Kawamuro:2018eky} and INTEGRAL~\cite{2014NIMPA.742...47U} offer larger sky coverage although with lower sensitivity. In this band, the \emph{eROSITA} all-sky survey~\cite{eROSITA:2020emt} will be key in providing a reference catalog against which flaring states of new or known extragalactic sources detected in neutrino ToO observations can be compared. Missions like \emph{SVOM}~\cite{10.1117/12.925171} (to be launched in 2023) will also perform rapid neutrino follow-ups, while more sensitive telescopes like \emph{Athena} will be used for detailed studies~\cite{Piro:2021oaa}. 

Correlation studies involving violent transients such as TDEs, GRBs and SNe with optical counterparts are already underway using observations from existing survey telescopes such as PanSTARRS~\cite{Pan-STARRS:2019szg}, DES~\cite{DES:2019hqa} and ZTF~\cite{Rauch:2019ntu}, and will be augmented in the coming years with the start of the LSST survey of the Vera Rubin Observatory~\cite{LSSTScience:2009jmu}. Wide multicolor optical surveys such as LSST and those conducted by space missions such as \emph{Euclid}~\cite{EUCLID:2011zbd}, SPHEREx~\cite{Dore:2018kgp} and the Roman Space Telescope~\cite{WFIRST:2018mpe} will also be critical to catalog other potential neutrino source classes such as star-forming galaxies~\cite{Lunardini:2019zcf}. 
In the microwave range, future observatories such as CMB-S4~\cite{Abazajian:2019eic} can provide photometric coverage of extragalactic sources such as radio-loud AGN and enable correlated searches for transients, while in radio upcoming facilities such as ngVLA~\cite{2018SPIE10700E..1OS} and ngEHT~\cite{Blackburn:2019bly} can provide detailed, time-dependent imaging of the jet and core regions of potential sources such as AGN. 

Gravitational wave detectors will also undergo important upgrades in the coming decade, starting with the addition of the KAGRA observatory to the existing LIGO and Virgo facilities~\cite{KAGRA:2013rdx} during their O4 run in 2022. 

\subsubsection{Interconnecting Multimessenger Facilities}
As the neutrino emission may be due to transient or bursting/flaring extragalactic sources, it is critical that follow-up observations are performed shortly after a neutrino signal is identified. This requires that neutrino telescopes operate realtime searches that can isolate potential astrophysical signals, either as individual high-energy neutrinos or spatial clusters of neutrino events over a given time scale, and a network over which these alerts can be broadcasted to the astronomical community for follow-up. IceCube sends out public realtime alerts for individual muon track and cascade events (singlets) of potential astrophysical origin, as well as private alerts to partner observatories for muon track multiplets over different time scales~\cite{2017realtimeIC}. Public alerts are communicated via the Gamma-ray Coordinates Network (GCN). ANTARES also operates singlet and doublet alert streams~\cite{2012AlertAN} while upcoming instruments such as KM3NeT~\cite{KM3NeT:2021dsa} and GVD~\cite{Baikal-GVD:2021ypx} are already developing their realtime programs.

The large number of realtime streams becoming available for current and future multimessenger instruments has called for an upgrade of the communication infrastructure used to broadcast alerts, especially given the enormous alert rates expected for facilities such as the Rubin observatory with an expected rate of 10 million alerts per night~\cite{ldm612}. Community alert brokers are currently under development to deliver this infrastructure for Rubin, while other efforts such as Scalable Cyberinfrastructure to support Multi-Messenger Astrophysics (SciMMA)~\cite{2020AAS...23510703B} are more targeted towards the multimessenger community. GCN is expected to be superseded in the coming years by the Time-domain Astronomy Coordination Hub (TACH~\cite{Sambruna:2021iki}), while networks like Astrophysical Multimessenger Observatory Network (AMON)~\cite{Smith:2012eu,AyalaSolares:2019iiy} are dedicated to realtime correlation studies of subthreshold streams from multimessenger observatories.

%% file: review.bbl
\hyphenation{Post-Script Sprin-ger}
\begin{thebibliography}{228}
\expandafter\ifx\csname natexlab\endcsname\relax\def\natexlab#1{#1}\fi

\bibitem{PhysRevLett.20.1205}
Davis R, Harmer DS, Hoffman KC.
\newblock \textit{Phys. Rev. Lett.} 20:1205 (1968)

\bibitem{PhysRevLett.20.1209}
Bahcall JN, Bahcall NA, Shaviv G.
\newblock \textit{Phys. Rev. Lett.} 20:1209 (1968)

\bibitem{PhysRevLett.58.1490}
Hirata K, et~al.
\newblock \textit{Phys. Rev. Lett.} 58:1490 (1987)

\bibitem{PhysRevLett.58.1494}
Bionta RM, et~al.
\newblock \textit{Phys. Rev. Lett.} 58:1494 (1987)

\bibitem{Aartsen:2013jdh}
Aartsen M, et~al.
\newblock \textit{Science} 342:1242856 (2013)

\bibitem{doi:10.1146/annurev-astro-081811-125539}
Haxton W, Hamish~Robertson R, Serenelli AM.
\newblock \textit{Annual Review of Astronomy and Astrophysics} 51:21 (2013)

\bibitem{doi:10.1146/annurev-nucl-102711-095006}
Scholberg K.
\newblock \textit{Annual Review of Nuclear and Particle Science} 62:81 (2012)

\bibitem{nusn_tamborra}
Tamborra I, Murase K.
\newblock \textit{Space Science Reviews} 214:31 (2018)

\bibitem{Markov:1960vja}
Markov MA. 1960.
\newblock In \textit{{Proceedings, 10th International Conference on High-Energy
  Physics (ICHEP 60): Rochester, NY, USA, 25 Aug - 1 Sep 1960}}

\bibitem{Reines:1960we}
Reines F.
\newblock \textit{Ann. Rev. Nucl. Part. Sci.} 10:1 (1960)

\bibitem{Greisen:1960wc}
Greisen K.
\newblock \textit{Ann. Rev. Nucl. Part. Sci.} 10:63 (1960)

\bibitem{RevModPhys.64.259}
Roberts A.
\newblock \textit{Rev. Mod. Phys.} 64:259 (1992)

\bibitem{2012EPJH...37..515S}
{Spiering} C.
\newblock \textit{European Physical Journal H} 37:515 (2012)

\bibitem{Collaboration:2011nsa}
Ageron M, et~al.
\newblock \textit{Nucl. Instrum. Meth.} A656:11 (2011)

\bibitem{Adrian-Martinez:2016fdl}
Adrian-Martinez S, et~al.
\newblock \textit{J. Phys.} G43:084001 (2016)

\bibitem{Sinopoulou:2021rgv}
Sinopoulou A, Coniglione R, Muller R, Tzamariudaki E.
\newblock \textit{PoS} ICRC2021:1134 (2021)

\bibitem{Avrorin:2019dli}
Avrorin AD, et~al. 2019.
\newblock In \textit{{36th International Cosmic Ray Conference (ICRC 2019)
  Madison, Wisconsin, USA, July 24-August 1, 2019}}

\bibitem{Baikal-GVD:2021zsq}
Belolaptikov I, et~al.
\newblock \textit{PoS} ICRC2021:002 (2021)

\bibitem{Aartsen:2016nxy}
Aartsen MG, et~al.
\newblock \textit{JINST} 12:P03012 (2017)

\bibitem{doi:10.1146/annurev-nucl-102313-025321}
Gaisser T, Halzen F.
\newblock \textit{Annual Review of Nuclear and Particle Science} 64:101 (2014)

\bibitem{2013NIMPA.700..188A}
{Abbasi} R, et~al.
\newblock \textit{Nuclear Instruments and Methods in Physics Research A}
  700:188 (2013)

\bibitem{Collaboration:2011ym}
Abbasi R, et~al.
\newblock \textit{Astropart. Phys.} 35:615 (2012)

\bibitem{2019inelasticity}
Aartsen M, et~al.
\newblock \textit{Physical Review D} 99 (2019)

\bibitem{PhysRev.118.316}
Glashow SL.
\newblock \textit{Phys. Rev.} 118:316 (1960)

\bibitem{Chirkin:2004hz}
Chirkin D, Rhode W  arXiv:hep-ph/0407075 [hep-ph] (2004)

\bibitem{doi:10.1146/annurev.nucl.50.1.679}
Learned JG, Mannheim K.
\newblock \textit{Annual Review of Nuclear and Particle Science} 50:679 (2000)

\bibitem{Albert:2018yoj}
Albert A, et~al.
\newblock \textit{Eur. Phys. J.} C78:1006 (2018)

\bibitem{Aartsen:2019epb}
Aartsen MG, et~al.
\newblock \textit{Astrophys. J.} 886:12 (2019)

\bibitem{2017showerrecoAntares}
Albert A, et~al.
\newblock \textit{The European Physical Journal C} 77 (2017)

\bibitem{2016crpp.book.....G}
{Gaisser} TK, {Engel} R, {Resconi} E (2016)

\bibitem{Enberg:2008te}
Enberg R, Reno MH, Sarcevic I.
\newblock \textit{Phys. Rev.} D78:043005 (2008)

\bibitem{Learned:1994wg}
Learned JG, Pakvasa S.
\newblock \textit{Astropart.Phys.} 3:267 (1995)

\bibitem{Aartsen:2013bka}
Aartsen MG, et~al.
\newblock \textit{Phys. Rev. Lett.} 111:021103 (2013)

\bibitem{PhysRevD.90.023009}
Gaisser TK, Jero K, Karle A, van Santen J.
\newblock \textit{Phys. Rev. D} 90:023009 (2014)

\bibitem{PhysRevD.104.022002}
Abbasi R, et~al.
\newblock \textit{Phys. Rev. D} 104:022002 (2021)

\bibitem{abbasi2021improved}
Abbasi R, et~al.  arXiv:2111.10299 [astro-ph.HE] (2021)

\bibitem{PhysRevLett.125.121104}
Aartsen MG, et~al.
\newblock \textit{Phys. Rev. Lett.} 125:121104 (2020)

\bibitem{Drury:1983zz}
Drury LO.
\newblock \textit{Rept. Prog. Phys.} 46:973 (1983)

\bibitem{Fermi-LAT:2014ryh}
Ackermann M, et~al.
\newblock \textit{Astrophys. J.} 799:86 (2015)

\bibitem{Murase:2018utn}
Murase K, Fukugita M.
\newblock \textit{Phys. Rev. D} 99:063012 (2019)

\bibitem{Murase:2016gly}
Murase K, Waxman E.
\newblock \textit{Phys. Rev.} D94:103006 (2016)

\bibitem{Yoshida:2020div}
Yoshida S, Murase K.
\newblock \textit{Phys. Rev. D} 102:083023 (2020)

\bibitem{Murase:2013rfa}
Murase K, Ahlers M, Lacki BC.
\newblock \textit{Phys.Rev.} D88:121301 (2013)

\bibitem{TheFermi-LAT:2015ykq}
Ackermann M, et~al.
\newblock \textit{Phys. Rev. Lett.} 116:151105 (2016)

\bibitem{Murase:2015xka}
Murase K, Guetta D, Ahlers M.
\newblock \textit{Phys. Rev. Lett.} 116:071101 (2016)

\bibitem{Capanema:2020rjj}
Capanema A, Esmaili A, Murase K.
\newblock \textit{Phys. Rev. D} 101:103012 (2020)

\bibitem{Capanema:2020oet}
Capanema A, Esmaili A, Serpico PD.
\newblock \textit{JCAP} 02:037 (2021)

\bibitem{IceCube:2016umi}
Aartsen MG, et~al.
\newblock \textit{Astrophys. J.} 833:3 (2016)

\bibitem{IceCube:2018fhm}
Aartsen MG, et~al.
\newblock \textit{Phys. Rev. D} 98:062003 (2018)

\bibitem{Aartsen:2019fau}
Aartsen MG, et~al.
\newblock \textit{Phys. Rev. Lett.} 124:051103 (2020)

\bibitem{Albert:2017ohr}
Albert A, et~al.
\newblock \textit{Phys. Rev.} D96:082001 (2017)

\bibitem{Aartsen:2020xpf}
Albert A, et~al.
\newblock \textit{The Astrophysical Journal} 892:92 (2020)

\bibitem{Fermi-LAT:2019yla}
Abdollahi S, et~al.
\newblock \textit{Astrophys. J. Suppl.} 247:33 (2020)

\bibitem{Fermi-LAT:2015bdd}
Ackermann M, et~al.
\newblock \textit{Astrophys. J.} 810:14 (2015)

\bibitem{Wakely:2007qpa}
Wakely SP, Horan D. 2007.
\newblock In \textit{{30th International Cosmic Ray Conference}}, vol.~3

\bibitem{2012AlertAN}
Ageron M, et~al.
\newblock \textit{Astroparticle Physics} 35:530–536 (2012)

\bibitem{2017realtimeIC}
Aartsen M, et~al.
\newblock \textit{Astroparticle Physics} 92:30–41 (2017)

\bibitem{Dornic:2019ag}
Dornic D, et~al.
\newblock \textit{PoS} ICRC2019:872 (2019)

\bibitem{Suvorova:2021ou}
Suvorova O, et~al.
\newblock \textit{PoS} ICRC2021:946 (2021)

\bibitem{2021FARIC}
Abbasi R, et~al.
\newblock \textit{The Astrophysical Journal} 910:4 (2021)

\bibitem{Aartsen:2018ywr}
Aartsen MG, et~al.
\newblock \textit{Eur. Phys. J.} C79:234 (2019)

\bibitem{2021allskyflare}
Abbasi R, et~al.
\newblock \textit{The Astrophysical Journal} 911:67 (2021)

\bibitem{Spiering:2019}
Spiering C  arXiv:1903.11481 [astro-ph.HE] (2019)

\bibitem{Blaufuss:2019fgv}
Blaufuss E, Kintscher T, Lu L, Tung CF.
\newblock \textit{PoS} ICRC2019:1021 (2020)

\bibitem{kopper2017grb}
Kopper C, Blaufuss E, Collaboration I, et~al.
\newblock \textit{Circular Service} :1 (2017)

\bibitem{IceCube:2018dnn}
Aartsen MG, et~al.
\newblock \textit{Science} 361:eaat1378 (2018)

\bibitem{doi:10.1126/science.aat2890}
Aartsen M, et~al.
\newblock \textit{Science} 361:147 (2018)

\bibitem{2015blazarANT}
Adrian-Martinez S, et~al.
\newblock \textit{JCAP} 12:014 (2015)

\bibitem{2017blazarIC}
Aartsen MG, et~al.
\newblock \textit{The Astrophysical Journal} 835:45 (2017)

\bibitem{Hooper:2018wyk}
Hooper D, Linden T, Vieregg A.
\newblock \textit{JCAP} 02:012 (2019)

\bibitem{Smith:2020oac}
Smith D, Hooper D, Vieregg A.
\newblock \textit{JCAP} 03:031 (2021)

\bibitem{2021stackingANT}
Albert A, et~al.
\newblock \textit{The Astrophysical Journal} 911:48 (2021)

\bibitem{2012Nature}
Abbasi R, et~al.
\newblock \textit{Nature} 484:351 (2012)

\bibitem{Li:2011ah}
Li Z.
\newblock \textit{Phys.Rev.} D85:027301 (2012)

\bibitem{Hummer:2011ms}
Hummer S, Baerwald P, Winter W.
\newblock \textit{Phys.Rev.Lett.} 108:231101 (2012)

\bibitem{He:2012tq}
He HN, et~al.
\newblock \textit{Astrophys.J.} 752:29 (2012)

\bibitem{2013ANTARESGRB}
Adrián-Martínez S, et~al.
\newblock \textit{Astron. Astrophys.} 559:A9 (2013)

\bibitem{2015ICGRB}
Aartsen MG, et~al.
\newblock \textit{The Astrophysical Journal} 805:L5 (2015)

\bibitem{Aartsen:2016qcr}
Aartsen MG, et~al.
\newblock \textit{Astrophys. J.} 824:115 (2016)

\bibitem{Adrian-Martinez:2016xij}
Adrián-Martínez S, et~al.
\newblock \textit{Eur. Phys. J.} C77:20 (2017)

\bibitem{Aartsen:2017wea}
Aartsen MG, et~al.
\newblock \textit{Astrophys. J.} 843:112 (2017)

\bibitem{2017ANTARESGRB}
Albert A, et~al.
\newblock \textit{Monthly Notices of the Royal Astronomical Society}
  469:906–915 (2017)

\bibitem{2020ANTARESGRB}
Albert A, et~al.
\newblock \textit{Monthly Notices of the Royal Astronomical Society}
  500:5614–5628 (2020)

\bibitem{Murase:2019tjj}
Murase K, Bartos I.
\newblock \textit{Ann. Rev. Nucl. Part. Sci.} 69:477 (2019)

\bibitem{IceCube:2015jsn}
Aartsen MG, et~al.
\newblock \textit{Astrophys. J.} 811:52 (2015)

\bibitem{Senno:2017vtd}
Senno N, Murase K, M\'esz\'aros P.
\newblock \textit{JCAP} 01:025 (2018)

\bibitem{Esmaili:2018wnv}
Esmaili A, Murase K.
\newblock \textit{JCAP} 12:008 (2018)

\bibitem{IceCube:2021oiv}
Necker J, et~al.
\newblock \textit{PoS} ICRC2021:1116 (2021)

\bibitem{2021TDE}
Stein R, et~al.
\newblock \textit{Nature Astronomy} 5:510–518 (2021)

\bibitem{2019TDEIC}
Stein R.
\newblock \textit{Proceedings of 36th International Cosmic Ray Conference —
  PoS(ICRC2019)}  (2019)

\bibitem{2021TDEANT}
Albert A, et~al.
\newblock \textit{The Astrophysical Journal} 920:50 (2021)

\bibitem{GBM:2017lvd}
Abbott BP, et~al.
\newblock \textit{Astrophys. J.} 848:L12 (2017)

\bibitem{ANTARES:2017bia}
Albert A, et~al.
\newblock \textit{Astrophys. J.} 850:L35 (2017)

\bibitem{Avrorin:2018fzl}
Avrorin AD, et~al.
\newblock \textit{JETP Lett.} 108:787 (2018)

\bibitem{Albert:2017obm}
Albert A, et~al.
\newblock \textit{Eur. Phys. J.} C77:911 (2017)

\bibitem{ANTARES:2017iky}
Albert A, et~al.
\newblock \textit{Phys. Rev.} D96:022005 (2017)

\bibitem{Adrian-Martinez:2016xgn}
Adrian-Martinez S, et~al.
\newblock \textit{Phys. Rev.} D93:122010 (2016)

\bibitem{Albert:2018jnn}
Albert A, et~al.
\newblock \textit{Astrophys. J.} 870:134 (2019)

\bibitem{2020ICGW}
Aartsen MG, et~al.
\newblock \textit{The Astrophysical Journal} 898:L10 (2020)

\bibitem{abbasi2021probing}
Abbasi R, et~al.
\newblock Probing neutrino emission at gev energies from compact binary mergers
  with the icecube neutrino observatory arXiv:2105.13160 [astro-ph.HE] (2021)

\bibitem{2020ANTARESGW}
Albert A, et~al.
\newblock \textit{The European Physical Journal C} 80 (2020)

\bibitem{2013ANTARESUHECR}
Adrián-Martínez S, et~al.
\newblock \textit{The Astrophysical Journal} 774:19 (2013)

\bibitem{Aartsen:2016ngq}
Aartsen MG, et~al.
\newblock \textit{Phys. Rev. Lett.} 117:241101 (2016), [Erratum: Phys. Rev.
  Lett.119,no.25,259902(2017)]

\bibitem{Mannheim:1995mm}
Mannheim K.
\newblock \textit{Astropart.Phys.} 3:295 (1995)

\bibitem{Aharonian:2000pv}
Aharonian FA.
\newblock \textit{New Astron.} 5:377 (2000)

\bibitem{Mucke:2000rn}
Mucke A, Protheroe RJ.
\newblock \textit{Astropart. Phys.} 15:121 (2001)

\bibitem{Murase:2015ndr}
Murase K arXiv:1511.01590 [astro-ph.HE] (2017).
\newblock {Active Galactic Nuclei as High-Energy Neutrino Sources}.
\newblock  15--31

\bibitem{Padovani:2019xcv}
Padovani P, et~al.
\newblock \textit{Mon. Not. Roy. Astron. Soc.} 484:L104 (2019)

\bibitem{Keivani:2018rnh}
Keivani A, et~al.
\newblock \textit{Astrophys. J.} 864:84 (2018)

\bibitem{MAGIC:2018sak}
Ansoldi S, et~al.
\newblock \textit{Astrophys. J. Lett.} 863:L10 (2018)

\bibitem{Gao:2018mnu}
Gao S, Fedynitch A, Winter W, Pohl M.
\newblock \textit{Nature Astron.} 3:88 (2019)

\bibitem{Cerruti:2018tmc}
Cerruti M, et~al.
\newblock \textit{Mon. Not. Roy. Astron. Soc.} 483:L12 (2019), [Erratum:
  Mon.Not.Roy.Astron.Soc. 502, L21--L22 (2021)]

\bibitem{Gasparyan:2021oad}
Gasparyan S, B\'egu\'e D, Sahakyan N.
\newblock \textit{Mon. Not. Roy. Astron. Soc.} 509:2102 (2021)

\bibitem{Murase:2018iyl}
Murase K, Oikonomou F, Petropoulou M.
\newblock \textit{Astrophys. J.} 865:124 (2018)

\bibitem{Rodrigues:2018tku}
Rodrigues X, et~al.
\newblock \textit{Astrophys. J. Lett.} 874:L29 (2019)

\bibitem{Reimer:2018vvw}
Reimer A, Boettcher M, Buson S.
\newblock \textit{Astrophys. J.} 881:46 (2019), [Erratum: Astrophys.J. 899, 168
  (2020)]

\bibitem{Petropoulou:2019zqp}
Petropoulou M, et~al.
\newblock \textit{Astrophys. J.} 891:115 (2020)

\bibitem{Zhang:2019htg}
Zhang BT, Petropoulou M, Murase K, Oikonomou F.
\newblock \textit{Astrophys. J.} 889:118 (2020)

\bibitem{Xue:2019txw}
Xue R, et~al.  arXiv:1908.10190 [astro-ph.HE] (2019)

\bibitem{Kadler:2016ygj}
Kadler M, et~al.
\newblock \textit{Nature Phys.} 12:807 (2016)

\bibitem{Giommi:2020viy}
Giommi P, et~al.
\newblock \textit{Astron. Astrophys.} 640:L4 (2020)

\bibitem{Petropoulou:2020pqh}
Petropoulou M, et~al.
\newblock \textit{Astrophys. J.} 899:113 (2020)

\bibitem{Rodrigues:2020fbu}
Rodrigues X, et~al.
\newblock \textit{Astrophys. J.} 912:54 (2021)

\bibitem{Oikonomou:2021akf}
Oikonomou F, et~al.
\newblock \textit{JCAP} 10:082 (2021)

\bibitem{Ber77}
Berezinsky VS.
\newblock \textit{{Proc. of Neutrinos-1977, Elbros.}}  (1977)

\bibitem{Eichler:1979yy}
Eichler D.
\newblock \textit{Astrophys. J.} 232:106 (1979)

\bibitem{PK83}
{Protheroe} RJ, {Kazanas} D.
\newblock \textit{Astrophys. J.} 265:620 (1983)

\bibitem{Stecker:1991vm}
Stecker FW, Done C, Salamon MH, Sommers P.
\newblock \textit{Phys.Rev.Lett.} 66:2697 (1991)

\bibitem{Murase:2019vdl}
Murase K, Kimura SS, Meszaros P.
\newblock \textit{Phys. Rev. Lett.} 125:011101 (2020)

\bibitem{Inoue:2019yfs}
Inoue Y, Khangulyan D, Doi A.
\newblock \textit{Astrophys. J. Lett.} 891:L33 (2020)

\bibitem{Anchordoqui:2021vms}
Anchordoqui LA, Krizmanic JF, Stecker FW.
\newblock \textit{PoS} ICRC2021:993 (2021)

\bibitem{Kheirandish:2021wkm}
Kheirandish A, Murase K, Kimura SS.
\newblock \textit{Astrophys. J.} 922:45 (2021)

\bibitem{2011Sci...333..203B}
{Bloom} JS, et~al.
\newblock \textit{Science} 333:203 (2011)

\bibitem{Wang:2015mmh}
Wang XY, Liu RY.
\newblock \textit{Phys. Rev.} D93:083005 (2016)

\bibitem{Senno:2016bso}
Senno N, Murase K, Meszaros P.
\newblock \textit{Astrophys. J.} 838:3 (2017)

\bibitem{Dai:2016gtz}
Dai L, Fang K.
\newblock \textit{Mon. Not. Roy. Astron. Soc.} 469:1354 (2017)

\bibitem{Lunardini:2016xwi}
Lunardini C, Winter W.
\newblock \textit{Phys. Rev.} D95:123001 (2017)

\bibitem{Murase:2020lnu}
Murase K, et~al.
\newblock \textit{Astrophys. J.} 902:108 (2020)

\bibitem{Hay21}
{Hayasaki} K.
\newblock \textit{Nature Astronomy} 5:436 (2021)

\bibitem{Reusch:2021ztx}
Reusch S, et~al.  arXiv:2111.09390 [astro-ph.HE] (2021)

\bibitem{Murase:2010cu}
Murase K, Thompson TA, Lacki BC, Beacom JF.
\newblock \textit{Phys.Rev.} D84:043003 (2011)

\bibitem{Katz:2011zx}
Katz B, Sapir N, Waxman E  arXiv:1106.1898 [astro-ph.HE] (2011)

\bibitem{Murase:2017pfe}
Murase K.
\newblock \textit{Phys. Rev.} D97:081301 (2018)

\bibitem{Petropoulou:2017ymv}
Petropoulou M, et~al.
\newblock \textit{Mon. Not. Roy. Astron. Soc.} 470:1881 (2017)

\bibitem{Waxman:1997ti}
Waxman E, Bahcall JN.
\newblock \textit{Phys.Rev.Lett.} 78:2292 (1997)

\bibitem{Murase:2005hy}
Murase K, Nagataki S.
\newblock \textit{Phys. Rev.} D73:063002 (2006)

\bibitem{Bahcall:2000sa}
Bahcall JN, Meszaros P.
\newblock \textit{Phys. Rev. Lett.} 85:1362 (2000)

\bibitem{Meszaros:2000fs}
Meszaros P, Rees MJ.
\newblock \textit{Astrophys. J. Lett.} 541:L5 (2000)

\bibitem{Murase:2013hh}
Murase K, Kashiyama K, M\'esz\'aros P.
\newblock \textit{Phys. Rev. Lett.} 111:131102 (2013)

\bibitem{Waxman:1999ai}
Waxman E, Bahcall JN.
\newblock \textit{Astrophys.J.} 541:707 (2000)

\bibitem{Dai:2000dj}
Dai Z, Lu T.
\newblock \textit{Astrophys.J.} 551:249 (2001)

\bibitem{Dermer:2000yd}
Dermer CD.
\newblock \textit{Astrophys.J.} 574:65 (2002)

\bibitem{Murase:2006dr}
Murase K, Nagataki S.
\newblock \textit{Phys. Rev. Lett.} 97:051101 (2006)

\bibitem{Guo:2019ljp}
Guo G, Qian YZ, Wu MR.
\newblock \textit{Astrophys. J.} 890:83 (2020)

\bibitem{Murase:2006mm}
Murase K, Ioka K, Nagataki S, Nakamura T.
\newblock \textit{Astrophys.J.} 651:L5 (2006)

\bibitem{Gupta:2006jm}
Gupta N, Zhang B.
\newblock \textit{Astropart. Phys.} 27:386 (2007)

\bibitem{Wang:2007ya}
Wang XY, Razzaque S, M\'esz\'aros P, Dai ZG.
\newblock \textit{Phys.Rev.} D76:083009 (2007)

\bibitem{Murase:2008mr}
Murase K, Ioka K, Nagataki S, Nakamura T.
\newblock \textit{Phys.Rev.} D78:023005 (2008)

\bibitem{Murase:2013ffa}
Murase K, Ioka K.
\newblock \textit{Phys.Rev.Lett.} 111:121102 (2013)

\bibitem{Senno:2015tsn}
Senno N, Murase K, Meszaros P.
\newblock \textit{Phys. Rev. D} 93:083003 (2016)

\bibitem{Tamborra:2015fzv}
Tamborra I, Ando S.
\newblock \textit{Phys. Rev. D} 93:053010 (2016)

\bibitem{Grichener:2021xeg}
Grichener A, Soker N.
\newblock \textit{Mon. Not. Roy. Astron. Soc.} 507:1651 (2021)

\bibitem{Murase:2009pg}
Murase K, M\'esz\'aros P, Zhang B.
\newblock \textit{Phys.Rev.} D79:103001 (2009)

\bibitem{Fang:2013vla}
Fang K, Kotera K, Murase K, Olinto AV.
\newblock \textit{Phys.Rev.} D90:103005 (2014)

\bibitem{Carpio:2020app}
Carpio J, Murase K.
\newblock \textit{Phys. Rev. D} 101:123002 (2020)

\bibitem{Kimura:2017kan}
Kimura SS, Murase K, M\'esz\'aros P, Kiuchi K.
\newblock \textit{Astrophys. J. Lett.} 848:L4 (2017)

\bibitem{Biehl:2017qen}
Biehl D, Heinze J, Winter W.
\newblock \textit{Mon. Not. Roy. Astron. Soc.} 476:1191 (2018)

\bibitem{Ahlers:2019fwz}
Ahlers M, Halser L.
\newblock \textit{Mon. Not. Roy. Astron. Soc.} 490:4935 (2019)

\bibitem{Kimura:2018vvz}
Kimura SS, et~al.
\newblock \textit{Phys. Rev. D} 98:043020 (2018)

\bibitem{Fang:2017tla}
Fang K, Metzger BD.
\newblock \textit{Astrophys. J.} 849:153 (2017)

\bibitem{Decoene:2019eux}
Decoene V, et~al.
\newblock \textit{JCAP} 04:045 (2020)

\bibitem{Murase:2008yt}
Murase K, Inoue S, Nagataki S.
\newblock \textit{Astrophys. J. Lett.} 689:L105 (2008)

\bibitem{Kotera:2009ms}
Kotera K, et~al.
\newblock \textit{Astrophys.J.} 707:370 (2009)

\bibitem{Zandanel:2014pva}
Zandanel F, Tamborra I, Gabici S, Ando S.
\newblock \textit{Astron. Astrophys.} 578:A32 (2015)

\bibitem{Fang:2016amf}
Fang K, Olinto AV.
\newblock \textit{Astrophys. J.} 828:37 (2016)

\bibitem{Fang:2017zjf}
Fang K, Murase K.
\newblock \textit{Nature Phys.} 14:396 (2018)

\bibitem{Hussain:2021dqp}
Hussain S, Alves~Batista R, de~Gouveia Dal~Pino EM, Dolag K.
\newblock \textit{Mon. Not. Roy. Astron. Soc.} 507:1762 (2021)

\bibitem{Loeb:2006tw}
Loeb A, Waxman E.
\newblock \textit{JCAP} 0605:003 (2006)

\bibitem{Thompson:2006np}
Thompson TA, Quataert E, Waxman E, Loeb A  arXiv:astro-ph/0608699 [astro-ph]
  (2006)

\bibitem{Tamborra:2014xia}
Tamborra I, Ando S, Murase K.
\newblock \textit{JCAP} 1409:043 (2014)

\bibitem{Chang:2014hua}
Chang XC, Wang XY.
\newblock \textit{Astrophys. J.} 793:131 (2014)

\bibitem{Senno:2015tra}
Senno N, et~al.
\newblock \textit{Astrophys. J.} 806:24 (2015)

\bibitem{Chakraborty:2015sta}
Chakraborty S, Izaguirre I.
\newblock \textit{Phys. Lett. B} 745:35 (2015)

\bibitem{Roth:2021lvk}
Roth MA, Krumholz MR, Crocker RM, Celli S.
\newblock \textit{Nature} 597:341 (2021)

\bibitem{Blanco:2021icw}
Blanco C, Linden T  arXiv:2104.03315 [astro-ph.HE] (2021)

\bibitem{IceCube-Gen2:2020qha}
Aartsen MG, et~al.
\newblock \textit{J. Phys. G} 48:060501 (2021)

\bibitem{P-ONE:2020ljt}
Agostini M, et~al.
\newblock \textit{Nature Astron.} 4:913 (2020)

\bibitem{Schumacher:2021hhm}
Schumacher LJ, et~al.
\newblock \textit{PoS} ICRC2021:1185 (2021)

\bibitem{KM3NeT:2018wnd}
Aiello S, et~al.
\newblock \textit{Astropart. Phys.} 111:100 (2019)

\bibitem{IceCube-Gen2:2021pdk}
Abbasi R, et~al.
\newblock \textit{PoS} ICRC2021:1186 (2021)

\bibitem{CTAConsortium:2013ofs}
Acharya BS, et~al.
\newblock \textit{Astropart. Phys.} 43:3 (2013)

\bibitem{2021arXiv210804512G}
{Gueta} O.
\newblock \textit{arXiv e-prints} :arXiv:2108.04512 (2021)

\bibitem{CTAConsortium:2017dvg}
Acharya BS, et~al.
\newblock WSP arXiv:1709.07997 [astro-ph.IM] (2018)

\bibitem{Sergijenko:2021lwn}
Sergijenko O, et~al.
\newblock \textit{PoS} ICRC2021:975 (2021)

\bibitem{VERITAS:2021mjg}
Acciari VA, et~al.
\newblock \textit{PoS} ICRC2021:960 (2021)

\bibitem{Abeysekara:2017mjj}
Abeysekara AU, et~al.
\newblock \textit{Astrophys. J.} 843:39 (2017)

\bibitem{Aharonian:2020iou}
Aharonian F, et~al.
\newblock \textit{Chin. Phys. C} 45:025002 (2021)

\bibitem{Hinton:2021rvp}
Hinton J.
\newblock \textit{PoS} ICRC2021:023 (2021)

\bibitem{APT:2021lhj}
Buckley JH, et~al.
\newblock \textit{PoS} ICRC2021:655 (2021)

\bibitem{NuSTAR}
Koglin J, et~al. 2009.
\newblock vol. 7437

\bibitem{Moiseev:2017mxg}
Moiseev A, et~al.
\newblock \textit{PoS} ICRC2017:798 (2018)

\bibitem{Fleischhack:2021mhc}
Fleischhack H.
\newblock \textit{PoS} ICRC2021:649 (2021)

\bibitem{Tomsick:2021wed}
Tomsick JA.
\newblock \textit{PoS} ICRC2021:652 (2021)

\bibitem{Evans:2015qia}
Evans PA, et~al.
\newblock \textit{Mon. Not. Roy. Astron. Soc.} 448:2210 (2015)

\bibitem{Kawamuro:2018eky}
Kawamuro T, et~al.
\newblock \textit{Astrophys. J. Suppl.} 238:32 (2018)

\bibitem{2014NIMPA.742...47U}
{Ubertini} P, {Bazzano} A.
\newblock \textit{Nuclear Instruments and Methods in Physics Research A} 742:47
  (2014)

\bibitem{eROSITA:2020emt}
Predehl P, et~al.
\newblock \textit{Astron. Astrophys.} 647:A1 (2021)

\bibitem{10.1117/12.925171}
Godet O, et~al. 2012.
\newblock In \textit{Space Telescopes and Instrumentation 2012: Ultraviolet to
  Gamma Ray}, eds. T~Takahashi, SS~Murray, JWA den Herder, vol. 8443.
  International Society for Optics and Photonics, SPIE

\bibitem{Piro:2021oaa}
Piro L, et~al.  arXiv:2110.15677 [astro-ph.HE] (2021)

\bibitem{Pan-STARRS:2019szg}
Kankare E, et~al.
\newblock \textit{Astron. Astrophys.} 626:A117 (2019)

\bibitem{DES:2019hqa}
Morgan R, et~al.
\newblock \textit{Astrophys. J.} 883:125 (2019)

\bibitem{Rauch:2019ntu}
Rauch L.
\newblock \textit{EPJ Web Conf.} 207:03001 (2019)

\bibitem{LSSTScience:2009jmu}
Abell PA, et~al.  arXiv:0912.0201 [astro-ph.IM] (2009)

\bibitem{EUCLID:2011zbd}
Laureijs R, et~al.  arXiv:1110.3193 [astro-ph.CO] (2011)

\bibitem{Dore:2018kgp}
Dor\'e O, et~al.  arXiv:1805.05489 [astro-ph.IM] (2018)

\bibitem{WFIRST:2018mpe}
Dor\'e O, et~al.  arXiv:1804.03628 [astro-ph.CO] (2018)

\bibitem{Lunardini:2019zcf}
Lunardini C, Vance GS, Emig KL, Windhorst RA.
\newblock \textit{JCAP} 10:073 (2019)

\bibitem{Abazajian:2019eic}
Abazajian K, et~al.  arXiv:1907.04473 [astro-ph.IM] (2019)

\bibitem{2018SPIE10700E..1OS}
{Selina} RJ, et~al. 2018.
\newblock In \textit{Ground-based and Airborne Telescopes VII}, vol. 10700 of
  \textit{Society of Photo-Optical Instrumentation Engineers (SPIE) Conference
  Series}

\bibitem{Blackburn:2019bly}
Blackburn L, et~al.  arXiv:1909.01411 [astro-ph.IM] (2019)

\bibitem{KAGRA:2013rdx}
Abbott BP, et~al.
\newblock \textit{Living Rev. Rel.} 21:3 (2018)

\bibitem{KM3NeT:2021dsa}
Assal W, et~al.
\newblock \textit{PoS} ICRC2021:941 (2021)

\bibitem{Baikal-GVD:2021ypx}
Suvorova OV, et~al.
\newblock \textit{PoS} ICRC2021:946 (2021)

\bibitem{ldm612}
Bellm E, et~al.
\newblock {LDM-612, Plans and Policies for LSST Alert Distribution}.
\newblock \url{https://ls.st/ldm-612} (2019)

\bibitem{2020AAS...23510703B}
{Brazier} A, et~al. 2020.
\newblock In \textit{American Astronomical Society Meeting Abstracts \#235},
  vol. 235 of \textit{American Astronomical Society Meeting Abstracts}

\bibitem{Sambruna:2021iki}
Sambruna RM, et~al.  arXiv:2109.10841 [astro-ph.IM] (2021)

\bibitem{Smith:2012eu}
Smith MWE, et~al.
\newblock \textit{Astropart. Phys.} 45:56 (2013)

\bibitem{AyalaSolares:2019iiy}
Ayala~Solares HA, et~al.
\newblock \textit{Astropart. Phys.} 114:68 (2020)

\end{thebibliography}
